\documentclass[a4paper,12pt]{article}

\usepackage{amssymb,amsmath,mathbbol,mathrsfs}
\usepackage[usenames,dvipsnames]{color}
\usepackage{hyperref}
\usepackage{xcolor}
\usepackage{stmaryrd}
\usepackage{authblk}
\usepackage{framed}
\usepackage{empheq}
\usepackage{slashed}
\usepackage{hyphenat}
\usepackage{cite}
\usepackage{apacite}

\usepackage[left=.7in,right=.7in,top=.75in,bottom=.75in]{geometry}                  

\usepackage{sectsty} 
\allsectionsfont{\sffamily\mdseries\upshape} 
\usepackage{tocloft}

\makeatletter
\renewenvironment{abstract}{%
    \if@twocolumn
      \section*{\abstractname}%
    \else 
      \begin{center}%
        {\bfseries\sffamily\abstractname\vspace{\z@}}
      \end{center}%
      \quotation
    \fi}
    {\if@twocolumn\else\endquotation\fi}
\makeatother

\numberwithin{equation}{section}
5

\hypersetup{
	colorlinks=true,         
	linkcolor=MidnightBlue,          
	citecolor=BrickRed,        
	urlcolor=MidnightBlue            
}

\newcommand{\be}{\begin{equation}}
\newcommand{\ee}{\end{equation}}

\newcommand{\Lie}{{L}}

\newcommand{\F}{{\Phi}}
\renewcommand{\d}{{\mathrm{d}}}
\newcommand{\D}{{\mathrm{D}}}

\newcommand{\SU}{{\mathrm{SU}}}
\newcommand{\U}{{\mathrm{U}}}

\newcommand{\G}{{\mathcal{G}}}
\newcommand{\A}{{\mathcal{A}}}

\newcommand{\pp}{{\partial}}

\newcommand{\fG}{{\mathrm{Lie}(\G)}}

\newcommand{\C}{{\mathbb{C}}}

\renewcommand{\bar}{\overline}

\newtheorem{defi}{Definition}
\newtheorem{theo}{Theorem}

\newcommand{\cint}{{\int\kern-.87em{<}}}
\newcommand{\sint}{{\int\kern-.75em{\sim}}}
\newcommand{\fint}{{\int\kern-1.00em{\int}}}

\newcommand{\bb}{\mathbb}

\newcommand{\tr}{\text{tr}}

\let\oldmarginpar\marginpar
\renewcommand\marginpar[1]{\oldmarginpar{\color{red}\raggedright\footnotesize #1}}

\newcommand{\old}{\color{red}}
\newcommand{\new}{\color{blue}}

\title{Holism as the empirical significance of symmetries\footnote{In honour of G. t' Hooft's 20th year Nobel prize celebration. }}
\author{Henrique Gomes \footnote{\href{mailto:gomes.ha@gmail.com}{gomes.ha@gmail.com}} \\\it University of Cambridge\\ \it Trinity College, CB2 1TQ, United Kingdom}
\begin{document}
\maketitle

\abstract{ 
Not all symmetries are on a par. For instance, within Newtonian mechanics, we seem to have a good grasp on the empirical significance  of  boosts, by applying it to subsystems. This is exemplified by the  thought experiment known as Galileo's ship:
the inertial state of motion of a ship is immaterial to how events unfold in the cabin, but is registered in the values of relational quantities such as the distance and velocity of the ship relative to the shore. 

But the significance of gauge symmetries seems less clear. For example, can   gauge transformations in Yang-Mills theory---taken as mere descriptive redundancy---exhibit a similar relational empirical significance as the boosts of Galileo's ship?  This question has been debated in the last fifteen years in philosophy of physics.

 I will argue that the answer is `yes', but only for a finite subset of gauge transformations, and under special conditions. Under those conditions, we can mathematically  identify empirical significance with a failure of supervenience: the state of the Universe is not uniquely determined by  the intrinsic state of its isolated subsystems. Empirical significance is therefore encoded in those relations between subsystems that stand apart from their intrinsic states.   }

\tableofcontents

\section{Introduction}

\subsection{Overview of the debate and my position within it}
In its broadest terms, a symmetry is a transformation of a system which preserves the values of a relevant (usually large) set of physical quantities. Of course, this broad idea is made precise in various different ways: for example as a map in the space of states, or on the set of quantities; and as a map that must respect the system's dynamics, e.g.  by mapping solutions to solutions or even by preserving the value of the Lagrangian functional on the states.

The broad idea is also associated with various debates.\footnote{See the essays in  \cite{BradingCastellani} and the references therein.} For example, should we say that a symmetry transformation applied to the whole universe cannot yield a different physical state of affairs?  And relatedly: should we prefer a reduced i.e. quotiented formalism, so that if presented with a state space $\mathcal{S}$ partitioned into the orbits of a group of symmetries $\G$ acting on $\mathcal{S}$, we prefer the reduced state space whose elements are the orbits, i.e. $[s]\in\mathcal{S}/\sim$ ? (where $s\sim s$ ($s, s'\in \mathcal{S}$) means that $s$ and $s'$ are related by a symmetry transformation: $s=\xi \cdot s$, for $\xi\in\G$ and $\cdot$ some action of $\G$ on $\mathcal{S}$,  and square brackets denote an entire equivalence class).

These ``defining features'' of symmetries are of central concern for one recent philosophical debate.   More specifically, the debate is about whether gauge symmetries can have a \textit{direct} empirical significance. 
 Of course, all hands agree that symmetries have various important empirical implications. The obvious examples come from the Noether theorems: the restrictions on the equations of motion entailed by Noether's second theorem, and the (approximately) conserved charges given by Noether's first theorem. In other words, symmetries imply the (extraordinary) facts that charges are conserved. 
 
  Accordingly, in this debate, such familiar implications are often called the `indirect' empirical significance of symmetries (IES); and through them, symmetries  carry immense explanatory power. 
 
  But some familiar symmetries of the whole Universe, such as velocity boosts in classical or relativistic mechanics (Galilean or Lorentz  transformations), have a \textit{direct} empirical significance  when applied solely to subsystems.  Thus Galileo's famous thought-experiment about the ship---that a process involving some set of  relevant physical quantities in the cabin below decks proceeds in exactly the same way  whether or not the ship is moving uniformly relative to the shore---shows that sub-system boosts have a direct, albeit relational, empirical significance. For though the inertial state of motion of the ship is undetectable to experimenters confined to the cabin, yet the entire system, composed of ship and sea\footnote{From now on,  I will prefer ``sea'' to ``shore''; this restriction eliminates the need to discuss  translations in addition to boosts \cite{MaudlinBuckets}, and places the two subsystems in direct contact, as in the case we will explore. } registers the difference between two such motions, namely in the different relative velocities of the ship to  the  water.\footnote{ 
As often is the case in physics, the characterization of DES used here may rely on certain approximations. Without such approximations,  the state in the cabin of Galileo's ship differs for different speeds and distances with respect to the shore, and no exact symmetries can be invoked. Nonetheless, approximate symmetries have consequences for theory construction (cf. \cite{Earman2019}). } Such examples rely on what are called `external symmetries', i.e. symmetries which shift spacetime points around.
  
 So the question arises: {\em Can other symmetries---especially gauge symmetries---have a similar direct empirical significance when applied to subsystems?}

For gauge symmetries are normally taken to encode descriptive redundancy: a view I will endorse. That is, they arise in a formalism that uses more  variables than there are physical degrees of freedom in the dynamical system described. (They are also internal: unlike a boost or spatial translation, they do \textit{not} shift spacetime points around).

This descriptive redundancy means that the natural answer to our question is `No'. For surely, while a ``freedom to redescribe'' may have some indirect empirical implications,\footnote{ By arriving at a local description to redescribe by weakening the global symmetries of the theory, the conserved charges implied by the global symmetries are required to  couple to fields in such a way that  conservation laws are dynamically respected  (this is the content of e.g. the Gauss law). } it could not have the content needed for a direct empirical significance, like the one illustrated by Galileo's ship. 
This `No' answer was   developed in detail by Brading and Brown \cite{BradingBrown} (henceforth BB) in response to  various discussions such as Kosso  \cite{Kosso}. They take themselves---I think rightly, in this respect---to be articulating the traditional or orthodox answer. 

The `Yes' answer has been argued for by Greaves and Wallace \cite{GreavesWallace} (henceforth GW), building on \cite{Healey2009}.  I will agree with some aspects of both BB's and GW's analysis of symmetries. But, unlike either GW  or BB, I will recast the topic to focus on gauge-\textit{invariant} information about---i.e. states of---regions. My own conclusion will be that only a finite subset of gauge-transformations, usually called `global' (but here called `rigid', cf. section \ref{sec:terminology}), can have DES. 

Before we can summarize the shape of the debate, we need a couple of definitions, which I will now informally sketch (exact definitions will appear in later sections).   First, the broad notion of `direct empirical significance' is a matter of the existence  of transformations of the universe possessing the  following two properties (articulated in this way by BB): \\
\indent \,(i) the transformation should lead to an empirically different  scenario, and\\
\indent  (ii) the transformation should be a symmetry of the subsystem in question (e.g. Galileo's ship).

 If such transformations exist,    the symmetries of the theory in question are said  to exhibit \textit{direct empirical significance} (henceforth `DES'). The empirical significance is to be witnessed by observers that lie outside the subsystem---it cannot be detected by those confined within it. Therefore, DES combines an inside and an outside perspective and, in this limited sense, acquires an epistemic dimension, or at least one that considers physical information as it is intrinsically accessible within a subsystem.
 
  BB argue that gauge symmetries cannot exhibit DES according to (i) and (ii), while GW argue that they can.  
I myself will argue for a `Yes' answer, but will approach the question in terms of gauge-{\em invariant} information.

 
By thus proceeding in terms of gauge-\textit{invariant} information, I will identify DES as defined by (i) and (ii) above with a failure of \textit{Global Supervenience on Subsystems} (GSS):\footnote{
 In the context of gauge systems under study here, a \textit{failure of global supervenience on subsystems}, is close in spirit to Myrvold's \textit{global patchy 
non-separability} \cite{Myrvold2010}, which he articulated for
the holonomy approach to gauge theories. But I refrain from adopting this
nomenclature because (i) I do not focus on holonomies, and (ii) it does not
apply to finite-dimensional systems like Galileo's ship. I will briefly comment again on this relation in footnote \ref{ftnt:Myrvold2} in Section \ref{sec:conc_sum}. \label{ftnt:Myrvold1} } this failure  is a form of holism---hence my title.  

I will show that relational DES occurs when the gauge-invariant global state fails to supervene on the
collection of intrinsic, gauge-invariant local states of the components of some arbitrary partition (of space or spacetime).   Here I should make it clear that I am not claiming \textit{ontological} priority for  the subsystems composing the whole. Subsystems don't exist ``before'' the whole.  The division of the Universe into subsystems is not mandatory, but it appears in item (ii): so we must consider what physical information is intrinsic to a subsystem when evaluating the direct empirical significance of symmetries. This is why, when environment and subsystems are on a par,   DES can be rephrased a matter of global supervenience.

GSS is upheld when the intrinsic physical states of those subsystems composing the whole uniquely determine the physical state of the whole, without the requirement for additional relational information about those subsystems. When GSS fails, there can either be many  physical states of the whole which are formed from the  same physical states of the individual subsystems---in which case one is missing some relational information---or there can be no valid states of the whole, in which case the subsystems states are incompatible. 

\subsection{External sophistication}\label{sec:soph}
If the standard notion of DES in (i) and (ii) is to be identified with a failure of GSS, we first need to develop a \textit{physically meaningful} notion of composition of those subsystems that possess descriptive redundancy.

 Thus, we are led to revisit one other important debate in the philosophy of gauge, already mentioned at the start of the previous section. Namely, 
 given a theory whose set of universe-descriptions---`states'---is partitioned by a group of symmetries, we can take one of two attitudes:\\
\indent (a)\,\textbf{Reduction}:-- try to write down a \textit{reduced} theory whose states correspond to the cells of the partition; or\\ 
 \indent (b)\,\textbf {Sophistication about symmetries}\footnote{Sophistication has long been advocated for diffeomorphisms and spacetime metrics (see \cite{Pooley_rel} and references therein). The nomenclature was originally used for sophisticated substantivalism: points of spacetime may have identity, but this identity comes only through the complex web of inter-relations between different fields  of the theory, and is in this way entirely dependent on the state. Some general features of  this position have more recently also been  suggested for gauge theories \cite{Caulton_gauge, Dewar2017, GomesStudies}. Dewar describes it thus: 
``Whereas a reduced theory converts a class of
symmetry-related models into a single model, sophistication converts a class
of symmetry-related models into a class of isomorphic models.''  Some philosophers (e.g. \cite[Sec. 4.2]{Healey_book} and \cite{Maudlin_response}) have resisted the analogy between the descriptive redundancy of the metric in gravitational theories and of the connection in gauge theories. \textit{Pace} these philosophers, I see no reason for their resistance, but that is a topic for another day. }:-- resist quotienting the given theory, but take two symmetry-related states to be isomorphic. \\

I'll advocate a third position (c), which applies only in the presence of subsystems.
Thus, for the entire universe, I unequivocally endorse reduction. In the same vein, I will also  assume that the theory in question  \textit{empirically} discerns two different states $s_1$ and $s_2$ of the universe \textit{if and only if} $[s_1]\neq[s_2]$. 
  
For subsystems, the question is more subtle, for there are two perspectives we can take: one from the inside, or intrinsic; and one from the outside, or extrinsic,  in accord with the definition of DES. If we are interested  in discriminating between (intrinsically) distinct physical possibilities, then surely those   states which cannot be (intrinsically) discriminated are to be counted as one, and so `reduction' still applies. 
 
 But,  when we combine the subsystem with the rest of the world, we are required to exploit subsystem symmetries in a real physical sense: as emphasized first by Rovelli \cite{RovelliGauge2013}, reduced representations of subsystems  cannot be straightforwardly coupled to each other.  For coupling, we need to keep  gauge-variant elements in the theory. In \cite{GomesStudies}, it was similarly argued that reduction should only be endorsed for the entire universe: coupling regional states may require a re-expression of the states as  particular gauge representatives of the physical states;\footnote{Rovelli focused on the coupling between different types of fields, or, in the finite-dimensional case, on the coupling of two different particle systems \cite{RovelliGauge2013}. In \cite{GomesStudies} I extended that requirement to the coupling of fields in  regions.  See also \cite{Dougherty2017} for a `stack-theoretic' argument  emphasizing the problems of reduction for the coupling of regions: his notion of separability  requires the preservation of gauge-related representations, to be kept as isomorphic but not identified. That is, I construe Dougherty's view as a defense of  position (b) motivated by the composition of subsystems (but using stack-theory). Also (implicitly) using a stack-theoretic approach, \cite{Teh_surplus} emphasize that gauge transformations are not just re-descriptions, but also parameterize the different ways in which regional field spaces can be composed into global field spaces.}  and thus, for regions, gauge-information should \emph{not} be entirely eliminated.

  Therefore, I have argued  that we should delineate a third attitude \cite{GomesStudies}:\\
\indent (c)\,\textbf{External sophistication and internal reduction of subsystem symmetries}:--Fix unique representations of the  intrinsic physical states of the subsystems (i.e. from the internal perspective),  but then allow these representations flexibility from an external perspective, as e.g. required for the smooth coupling of the states of subsystems.  \\

  Take the example used in  \cite{RovelliGauge2013}: a non-relativistic classical system of N particles with translational invariance. From the intrinsic perspective of the subsystem, one could eliminate redundancy by taking the inter-particle distances as a new, autonomous set of coordinates, but this would leave no `handle' for other subsystems to couple to. The joint system of autonomous coordinates for two such sets of particles (say $N_1$ `red' and $N_2$ `green') cannot express different ways of composing the subsystems---whether the center of mass of the `reds'  are five or ten feet away from the center of mass of the `greens' along some direction does not register in these variables. From a degree of freedom count, we have clearly gone overboard: we have eliminated six degrees of freedom of the joint system---the position of the center of mass of `reds' (three degrees of freedom) and the position of center of mass of `greens' (three degrees of freedom)---when only  three were eliminable: the position of center of mass of $\{$reds$\}\cup\{$ greens$\}$).  
  
  On the other hand, fixing the isolated subsystem's coordinates by reference to its center of mass, while leaving the center of mass embedded in Euclidean space, still affords us enough flexibility to characterize both the subsystems intrinsic degrees of freedom and a rigid subsystem translation with respect to another subsystem. This is a very simple example of a  ``covariant gauge-fixing'', and in practice, it is how we implement option (c).

In my analysis of DES in the context of holism, the flexibility allowed by option (c), `External sophistication' for short, is  employed for melding the subsystems' physical content into  the physical content of the joint state.\footnote{\cite{Teh_surplus} accept something like position (c). The main difference  is  that they don't impose a unique representative on regions (they just attach the entire groupoid of fields to each region). In other words,  they are not interested in uniquely and explicitly parametrizing the physical content of each region, employing instead a ``stack-theoretic" picture.}  Option (c) allows us to have our cake and eat it too: we can both parametrize the intrinsic physical possibilities of the subsystems in a  one-to-one manner, and yet keep track of those degrees of freedom that would be redundant from the intrinsic point of view but which must be retained, to  be pressed into service for composing subsystems \cite{GomesStudies}.

\subsection{Non-technical summary}

Note  that the difference between the two  scenarios above---between composing the reduced subsystems and composing the covariantly gauge-fixed subsystems---involves \textit{the space of possibilities} of the subsystems. The mismatch  is only apparent when we consider all the possible intrinsic physical states of the subsystems, say, $n$ in number,  and the physical states of the universe that are  compatible with each $n$-tuple of physical subsystem states. That is, it is only apparent in the same context of GSS, or lack thereof.

 Indeed,  a given  state of the entire universe will be taken to uniquely fix the physical subsystem states that constitute it, as well as any extra relational information between them. Given first the state of the universe, there is  no question about composition: only of \textit{de}composition, which is straightforward.  But GSS is not about decomposition, and neither is DES, even though this confusion permeates the debate, as we will briefly see in Section \ref{sec:GW_BB}. To avoid this confusion, it is best to start with the information that is intrinsically available to the subsystems and assess which universal physical states can emerge from composition. 

To be clear, in this paper, I will only countenance the case where a given physically allowed state of the universe  decomposes into physically allowed states of its subsystems. I know of no examples violating this assumption. Therefore,   we will focus on the more interesting of the two cases of failure of GS: namely, the one where, given just the intrinsic physical states of the subsystems, there are physically distinct possibilities to join these states into a state of the Universe. That is, the relation between states of the Universe and states of its subsystems are many-to-one, because there is relevant relational information that cannot be registered intrinsically within each subsystem. In these cases, we will say there is \textit{Global Non-Supervenience on Subsystems} (GNSS). 

 Schematically: if the subsystems are `sea' and `ship', and there are  equivalence relations, $\sim$,  applicable to states of  subsystems and of the whole, and  given  the physical (i.e. ``gauge-invariant'') states  $[s_{\text{\tiny sea}}], [s_{\text{\tiny ship}}]$ and $[s_{\text{\tiny sea and ship}}]$, there is a many-to-one relation, encoded by the set $I$:
\be\label{eq:seaship}[s_{\text{\tiny sea and ship}}]_{(i)}=[s_{\text{\tiny sea}}]\cup_{(i)}  [s_{\text{\tiny ship}}],\qquad i\in I=\text{Boosts}\ltimes\text{Euclidean},
\ee
and $[s_{\text{\tiny{sea and ship}}}]_{(i)}=[s_{\text{\tiny sea and ship}}]_{(i')}$ if and only if $i=i'$. In this case, the set $I$ that parameterizes the many-to-one relation is  the (inhomogeneous) Galilean group (which is a semi-direct product ($\ltimes$) of boosts and the group of translations and rotations).

 One can thus see that the variety of states of the whole is  \textit{not} encoded  in either subsystem:\footnote{Agreed, several approximations must be in place for this statement: for surely,  with the right equipment (such as a window),  the person within the cabin \textit{could} discern movement of the ship from within,  and different movements of the ship could create different sorts of eddies and turbulence in the sea. This sort of idealization is ubiquitous in physics, and generally unproblematic. This is echoed in \cite[p. 52, footnote 7]{GreavesWallace}: `` ‘Dynamical isolation’ is of course approximate in practice, as no proper subsystem can in
practice be perfectly isolated. It is also relative to the observation capabilities of relevant observers
(if Galileo had included a GPS tracker in his list of the cabin’s accoutrements, things
would have been rather different).'' See also \cite{Earman2019} for the utility of  approximate symmetries in physics. \label{ftnt:approx} } it is encoded in the \textit{relations}  between the two subsystems, as denoted by $\cup_{(i)}$.  

In certain situations, such as in Galileo's ship, there is remarkable order to this variety of physical states of the whole, an order also encoded in the structure of $I$. Namely, each element of this variety  can be transformed into another by  a subsystem symmetry which does not extend beyond the boundary of the subsystem. In other words, $I$ carries the structure of finite-dimensional symmetry group of the subsystem. For Galileo's ship, these are the Galilean transformations;  and for the gauge theory, we will see that they are (sub)groups of the Lie group characterizing the theory. 
In this manner, DES  becomes a matter of GNSS.

 In sum: technicalities apart, 
  my main claim is that both Galilean boost symmetry for particle systems and gauge symmetry for certain field theories carry direct empirical significance  through GNSS. This holism is \textit{empirically significant}, since it registers physical---i.e. gauge-invariant---differences in the entire system and we take such differences to lead to empirically distinguishable universes. Moreover, the implied under-determination of the physical state of the whole universe by the physical state of its subsystems is encoded in a \textit{subsystem symmetry}, but only  as seen from the `outside perspective'; again in accord with the above construal of DES.


\section{Direct empirical significance}\label{sec:DES}
I start, in  section \ref{sec:DES_uni}, by  construing DES in terms of properties of  transformations of the Universe. This description of DES immediately runs into some cumbersome notation when applied to gauge theories. Therefore, in section \ref{sec:terminology} I introduce a new terminology which better distinguishes the relevant categories of transformations. Then, having got the right nomenclature for addressing DES in the context of gauge theories, in section  \ref{sec:DES_debate} I apply it to re-express the debate in these better terms.  In section \ref{sec:intro_variety}, I then proceed to offer an appetizer of my criticism of BB and GW's construals of DES, and supplant those construals with my own. Thus here I  describe the relation between GNSS and DES. 


\subsection{DES as a transformation of the universe}\label{sec:DES_uni}
BB frame the  definition of DES in terms of two conditions.  
 First, a transformation cannot be a symmetry of the entire universe, otherwise it would not have any direct empirical significance. But second, it needs to act as a symmetry for subsystems, otherwise the transformation in question could hardly be called a symmetry. Thus
Brading and Brown  define: 
\begin{defi}[Direct Empirical Significance (DES) as a transformation]\label{def:DES}
A symmetry has  direct empirical significance if it is specified by a  transformation that satisfies the following conditions:\footnote{Both Teh \cite{Teh_emp} and GW add a condition of dynamical isolation between the two subsystems. Teh takes this to justify an asymptotic treatment for the subsystem in question. We won't need to make this isolation condition explicit: it emerges from the criteria. }
\begin{enumerate}
\item \textbf{Transformation Condition}: the transformation  must yield an empirically different scenario. In our words: the transformation in question is not a symmetry of the world as a whole. 
\item \textbf{Subsystem Symmetry Condition}: The evolution of the untransformed and transformed {subsystems} must be empirically indistinguishable from the interior point of view. In our words:  the transformation should  count as a symmetry when restricted to the subsystems composing the entire system.This is the subsystem symmetry with DES.\footnote{ Although BB distinguish a subsystem and its environment, and thus have only the singular `subsystem' in their definition,  it is customary to focus on \textit{relational} DES, i.e. DES with the environment  taken on a par with the subsystem in question. 
 Thereby, the environment is taken as  just one more subsystem,  and transformations of the environment are to be considered just as much as  transformations of the subsystem.  In particular, this demotion of the environment to subsystem status  means that one  excludes the exclusively external relations of the environment from its  state. A non-relational definition of DES  would not require  condition $\mathit{2}$ to apply to the environment (cf. footnote \ref{ftnt:principled}), that is, it would include transformations that are \emph{not} symmetries of the intrinsic state of the environment. The standard argument against non-relational DES is that only relational DES has a principled connection between a subsystem symmetry and physically distinct universes.  To glean the difficulty with non-relational DES, consider  the following example:  in the Galilieo ship scenario, imagine a transformation that arbitrarily  changes the configuration of the interior of the beach---taken as part of the environment---but otherwise keeps the subsystem-intrinsic and all other relational  information untouched. If definition \ref{def:DES} did not require the physical state of the environment to be preserved, such a transformation would be included and therefore correspond to a symmetry with  DES. But an arbitrarily changing beach has little to do with symmetries. See also footnote \ref{ftnt:truncation}. \label{ftnt:non_rel_DES}}

\end{enumerate}
\end{defi}

Thus in the example of Galileo's ship, the entire system---both ship and sea---is in different states if the ship is heading  through calm waters towards the North-West at 10km/h or towards the South at 20km/h. The entire system thus satisfies the first condition (\textit{Transformation}). Nonetheless, \textit{inside the cabin}, you would not be able to distinguish the two scenarios (cf. footnote \ref{ftnt:approx}): so the subsystem  satisfies the second condition.

As is clear from the above, the empirical significance cannot be witnessed by observers within the subsystem and, in that way, DES
combines an inside and an outside perspective: inside it considers intrinsic physical information
that is accessible within the subsystem; outside it considers an overall change in
the state of the universe.

\subsection{Two distinctions}\label{sec:terminology}
 At this point in the discussion, standard terminology gets in the way of clarity. When used in conjunction with subsystem-Universe distinctions, the words `local' and `global' acquire other possible meanings, and may pull intuition in different directions (see p. 648 of \cite{BradingBrown}). 
   Therefore, it is useful to introduce a nomenclature that distinguishes these meanings. 
\begin{itemize}
\item \textit{Universal}: A universal transformation is one that applies to the world as a whole. 
 The set of universal transformations may depend on an infinite or finite number of parameters. 
\item \textit{Regional}: A regional transformation is one that applies only to a subsystem of the world.\footnote{As in our discussion hitherto, this is often called a `subsystem symmetry'  \cite{GreavesWallace, Teh_emp}, but here I employ this alternative nomenclature because my interest will be solely in subsystems formed by restricting to a spacetime region. } 
The set of regional symmetry transformations may also depend on an infinite or finite number of paramaters. 
\item \textit{Malleable (aka `local')}:  A malleable transformation depends on an infinite number of parameters: e.g. it is specified by an arbitrary smooth function over a given manifold or region of a manifold. Here, the usual label is `local'. But using `local' invites confusion with the above category, `regional'. I will therefore prefer the term `malleable'. A malleable symmetry transformation can be either regional or universal. 
\item \textit{Rigid (aka `global')}:   A `rigid' symmetry transformation depends only on a finite number of parameters.  This is to be contrasted with \textit{malleable}. The usual label is `global'. But again, this term invites confusion, namely with the above category, `universal'. So I will  prefer the term `rigid'. A rigid symmetry transformation can be either regional or universal. 
\end{itemize}
Therefore a symmetry transformation may lie in any of the following four combinations of the above categories: regional and rigid, regional and malleable, universal and malleable, or universal and rigid.\footnote{Teh \cite{Teh_emp} labels the transformations with underlining: as \underline{\textit{local}} (meaning regional), and \textit{local} (meaning malleable), and   \underline{\textit{global}} (meaning universal), and \textit{global} (meaning rigid); but I feel this underlining also invites confusion. }

Regional transformations are under-studied in the physics literature, but are known to hide many surprises: see e.g. \cite{ ReggeTeitelboim1974, Balachandran:1994up, DonnellyFreidel, GomesHopfRiello, GomesStudies}.
 As to universal transformations,  the rigid ones are familiar; they are associated with the standard treatments of Noether's first theorem, and thus correspond to conserved charges \cite{Noether, Noether3, BradingBrown_Noether, Butterfield_symp, Noether2}. The malleable universal transformations are associated to constraints, or relations between  the equations of motion (such as the Hamiltonian constraints or the Bianchi identities of general relativity and the Gauss constraint in electromagnetism).\footnote{ In the Hamiltonian treatment, the symmetries are represented as flows in the constrained phase space, with orbits being the  manifold to which the (integrable) flows are tangent (see \cite{Earman_ode} for a celebration of the virtues of the Hamiltonian treatment). In the Lagrangian treatment, symmetries are represented as orbits in configuration space \cite{Lee:1990nz}. A powerful formalism which lies in between the Hamiltonian and Lagrangian  is the \textit{covariant symplectic formalism} \cite{Lee:1990nz, WittenCrnkovic}. It is most useful in discussing canonical (or Hamiltonian) features of a system while retaining easy access to spacetime covariance.} 
But since these constraints guarantee charge conservation as well, we note that it was argued in  \cite{GomesNoether}---in the nomenclature introduced above---that the real power of the Noether theorems arises from weakening a rigid symmetry to a malleable one, which thereby enforces  compatibility between  charge conservation and the dynamics of the fields that the charges couple to.

\subsection{The debate re-expressed}\label{sec:DES_debate} 
Using this nomenclature, we can  re-express Definition \ref{def:DES} and better address the subtleties of applying it to gauge symmetry. Thus Definition \ref{def:DES} says that DES arises if there are transformations that are not universal symmetries and yet  whose restrictions are regional symmetries. The question is which, if any, regional malleable symmetry can be obtained in this way,  and thus be awarded DES. 
 
  Finite-dimensional theories, i.e. ones which do not involve fields, generally only have rigid symmetries (such as translations, etc.). 
In those cases, the strictly regional symmetries can give different values for appropriate physical quantities, viz. relational quantities relating the transformed subsystem to the rest of the universe.   This is of course what Galileo's ship illustrates.  In this case, the clear distinction between universal and regional rigid symmetries is illustrated in an uncontroversial case of DES. 

 But the situation for malleable symmetries seems different.  In certain examples,  the  generators of malleable symmetries are  spacetime vector fields; in others, they are   Lie-algebra-valued scalar fields,  acting on an internal space over each  spacetime point. 
In any case,  it is easy to imagine a malleable symmetry acting on a region of spacetime and not on another, which can thereby serve as the environment, or reference system. In this case,  the malleable transformation should smoothly tend to the identity at the boundary between  the regions, lest it create discontinuities in the fields. But then, it seems  we could suitably extend any such regional malleable  symmetry to the rest of the  Universe simply with the identity transformation. The conjunction of the two regional transformations---one that tends to the identity at the boundary and the other the identity on the rest of the universe---would be  a universal malleable symmetry,  and thus could not have DES.

According to BB \cite{BradingBrown},  this is precisely the case: the environment can be assumed to be untransformed, and therefore, to avoid discontinuities and due to their malleability, gauge symmetries would violate Definition \ref{def:DES}'s first condition (\textit{Transformation}), i.e. the requirement of an empirically distinguishable scenario, and so would have no empirical significance in the way that regional rigid symmetries do. 
As BB write:
\begin{quote}
``Thus, a transformation applied to one subsystem will involve the other subsystem, even if only because the transformation of the gauge field goes smoothly to the identity. In conclusion, there can be no analogue of the Galilean ship experiment for local gauge transformations, and therefore local gauge symmetry has only indirect empirical significance (being a property of the equations of motion).'' (p. 657)\end{quote}


GW articulate DES for gauge theory in a manner that fosters DES for gauge symmetries. They focus on subsystems as given by regions, and thereby identify transformations possessing properties  $\mathit{1}$ and $\mathit{2}$ of Definition \ref{def:DES} by first formulating the putative effects of such transformations on the gauge fields in these regions. 

In particular, they focus their attention on  \textit{relational} DES. That means they consider the environment to be on a par with the subsystem in question (cf. footnote \ref{ftnt:non_rel_DES}). Thus, in particular, the transformation  in Definition \ref{def:DES} must obey  property $\mathit{2}$---it must also be a symmetry of the environment of the subsystem.\footnote{While GW do allow for the larger, non-strictly relational quotient group---of all subsystem symmetries quotiented by the interior ones, where condition $\mathit{2}$ need not apply to the environment, or the complement of the subsystem---they do not investigate this overly general definition of DES. For,  in their nomenclature, there could be no `principled connection' between an element of the wider group and empirical significance \cite[p.86,87]{GreavesWallace}. See footnotes \ref{ftnt:non_rel_DES} and \ref{ftnt:truncation} here for more on the `principled connection' and the treatment of the environment as solely a reference and not a subsystem; and  \cite{Teh_abandon} for one possible interpretation of the term `principled connection' for non-relational DES.  \label{ftnt:principled}} 
 In this case we can diagnose DES as originating in  the relations between the subsystems (and, ultimately, as I will show,  in a failure of supervenience of the global state on the intrinsic states of the subsystems).

More precisely,  for a given subsystem state $s$, they claim the relational DES transformations are in 1-1 correspondence with the following quotient between two  groups of transformations:
\be\label{eq:DESGW_final}\G^{\text{\tiny{GW}}}_{\text{\tiny DES}}(s)\simeq \mathcal{G}(s{}_{|\pp})/\mathcal{G}_{\text{Id}},\ee
 where $\mathcal{G}(s{}_{|\pp})$ are the gauge transformations of the region which preserve the state $s$ \textit{at the boundary} of the region, and $\mathcal{G}_{\text{Id}}$ are the gauge transformations of the region which are the identity at the boundary. Here the equivalence class between two transformations ${\xi}, {\xi'}\in  \mathcal{G}(s{}_{|\pp})$ is taken as ${\xi}\sim {\xi'}$ iff ${\xi}'=\eta\xi$ for some $\eta\in \mathcal{G}_{\text{Id}}$.\footnote{A group is just a set closed under an associative invertible binary operation. That is, if  ${\xi}, {\xi'}\in \G$, then ${\xi}{\xi'}\in \G$, $(\xi{\xi'}){\xi'}'=\xi({\xi'}{\xi'}')$ and for all ${\xi}\in \G$ there exists a ${\xi}^{-1}$ such that ${\xi}^{-1}\xi=$Id, where ${\xi}$Id$=\xi$  (here we will not need to distinguish left and right inverses). The quotient is well-defined for normal subgroups: namely, $\G/\mathcal{H}$ is well-defined as a group if given $\eta\in \mathcal{H}$, ${\xi}\eta\xi^{-1}\in \mathcal{H}$ for all ${\xi}\in \G$. This holds in the example above. } The rough idea is that even if certain transformations would not preserve all possible states at the boundary, they will preserve \emph{some} of those states.\footnote{One could think of it as follows: some state $s$  might be 'periodic' in what it says about the boundary; so that a transformation with the {\em same} period at the boundary will fix the state $s$ at the boundary even though the transformation is not the identity there (i.e does {\em not} fix all states $s'$ at the boundary. That is: when the transformation is restricted to boundary, it  does not fix each state thus restricted to boundary).
\label{ftnt:periodic}} So, $\mathcal{G}_{\text{Id}}$ is a subgroup of $\mathcal{G}(s{}_{|\pp})$, and one would like to `factorize out' from those transformations those that would preserve \emph{all} states at the boundary (i.e. the boundary-identity transformations): so one defines the quotient group $\mathcal{G}(s{}_{|\pp})/\mathcal{G}_{\text{Id}}$. 

Several assumptions go into the results of GW and BB, and we will challenge some of these once we have described our own results (see Section \ref{sec:GW_BB}). 


\subsection{My own position in the debate:  rigid variety}\label{sec:intro_variety}
Overall,  I will argue for a position not considered by either the GW or the BB camp: an appropriate selection of rigid regional symmetries---but not all the malleable ones!---can retain direct empirical significance (DES) in both the finite-dimensional case and in the field-theoretic case. In very specific circumstances, and according to a precise method, the rigid symmetries will be identified among the malleable ones---they are the ones that leave the gauge potential invariant but which shift the matter fields, and they will be the only ones lying in the kernel of a (configuration-dependent, in the non-Abelian case) elliptic differential operator---and they will be transformation with DES  as per Definition \ref{def:DES}.



 \paragraph*{Rigid variety and relational DES}\label{sec:under_intro} 
   In the following, to make matters concrete, the field-theories  I will focus on are general classical Yang-Mills theories in the presence of matter. The spacetime fields in question will be the standard, smooth gauge potentials, $A$, and charged scalars, $\psi$, valued in the appropriate vector spaces, which I discuss in more detail below, in Section \ref{sec:notation}. I denote the doublet of these two fields by: $\varphi=(A,\psi)$, and the space of such doublets by $\Phi\ni \varphi$.  The subsystems will consist of regions in the manifold, to which the fields get restricted. Thus the subsystems in questions are \textit{regional}, and thus we refer to \textit{Global Supervenience on Regions} (to avoid unnecessarily detailed acronyms, we will denote these also by GSS). This initial set-up  is  standard in the debate about the DES of gauge symmetries, and is applicable to all of the approaches considered here. 
   
   DES
combines an inside and an outside perspective: inside it considers physical information
that is intrinsically accessible to subsystems; outside it considers an overall change in
the state of the universe.
Definition \ref{def:DES} can be formulated as a failure of GSS because its requirement $\mathit{2}$ is about  information that is \textit{intrinsically} available to a subsystem. Therefore, for relational DES,  a transformation that does not change the intrinsic physical states of its subsystems and yet changes the physical state of the Universe must be changing the relations between the subsystems. Therefore, for DES to exist, there must be a physical variety of universes which are made up from the same (physically) subsystem states. 
   
   Accordingly, within this context of Yang-Mills fields, I define:\footnote{For the relation to Myrvold's ``patchy separability'' see footnotes \ref{ftnt:Myrvold1} and \ref{ftnt:Myrvold2}.}
\begin{defi}[GSS]\label{def:reg_ins}
Given a manifold $\Sigma$, that is decomposed as $\Sigma=\Sigma_+\cup\Sigma_-$, along the boundary $\pp \Sigma_\pm=\Sigma_+\cap\Sigma_-=:S$; given a universal field supported on $\varphi$ and  the regional   fields,  $\varphi^\pm$ supported on $\Sigma_\pm$,  \textbf{GSS} holds just in case the joint \textit{gauge-invariant contents} of  $\varphi^\pm$ is compatible with a unique gauge-invariant content of $\varphi$. That is,  the doublet of regional physical (i.e. gauge-invariant) states $([\varphi_+], [\varphi_-])$ uniquely  determines a valid physical state $[\varphi]$ for the field over the entire manifold $\Sigma$. 
\end{defi}
 More formally, we label each legitimate/physically possible composition of the two given regional states to form a physically possible universal state  by $i$, with $i$ belonging to some index set $I$, which can depend on the component states. Thus, formally,  failure of GSS corresponds to the set $I$ having either none or more than one element (if $I=\emptyset$, the two states are incompatible). 
 
 As in the more general case of subsystems (as opposed to regions), I will not countenance the possibility that the relation between universe and regional physical states is one-to-many (which would amount to the universe state being a coarse-graining of the conjunction of regional states), nor that there are valid physical states of the universe whose restrictions to regions are not themselves physically valid (I can see no plausible scenario in which that occurs). 
 
 Here we will focus on the case in which $I$ has many elements, and, as in the more general nomenclature for subsystems, we will call this  \textit{Global Non-Supervenience on Regions} (GNSS).  Thus, denoting the gauge-equivalence class by square brackets,  indicating the $i$th composition of states by $\cup_{(i)}$ and the resulting universal physical state as $[\varphi_{(i)}]$,\footnote{Even though each $i$ will represent a different physical state, we avoid putting the subscript outside of the equivalence class, because there is only one equivalence class of $\varphi$;  it can have no indexing. } we write 
  \be\label{eq:reg_ins} [\varphi_{(i)}]=[\varphi^+]\cup^S_i [\varphi^-],\quad i\in I\quad\text{with}\quad  [\varphi]_{(i)}\neq  [\varphi]_{(i')}\quad\text{iff}\quad i\neq i'\ee
 or, in terms of the standard Yang-Mills potential and matter fields:
 \be\label{eq:2.3}[A,\psi]^{(i)}=[A^+,\psi^+]\cup^S_i [A^-,\psi^-],\qquad i\in I,\ee
So here $i$ is neither a spacetime index nor necessarily related to a Lie-algebra index: it is just an element of an index set $I([\varphi^\pm])$---whose dependence  on the given pair $[\varphi^\pm]$ will be henceforth omitted---and $\cup^S_i$ represents the $i$-th valid \textit{gluing}, i.e. composition,  of the two gauge-invariant data $[\varphi^\pm]$ along $S$. The global fields $\varphi$ are in the same class of differentiability as the regional ones (albeit the latter will lie on manifolds with boundary).

  If $I$ is empty  there is \textit{no} possible gluing, i.e. the regional gauge-invariant states are incompatible and cannot conjoin into a universal physical state (\textit{regional incompatibility}). If $I$ has a single element, the gluing is unique, and then  there is  GSS. If otherwise, i.e. if $I$ has more than one element, the universal physical state is undetermined just by the regional physical states: more information about relations between the subsystems is needed,  and  there is {GNSS}. In this case, we will also say the universal state possesses \textit{residual variety}.

 Note that as it stands Definition \ref{def:reg_ins} is in line with both the Galileo's ship analogy and with the idea of gauge transformations as mere re-description. The analogy  states that, in general, the physical states $[\varphi_{\mbox{\tiny sea}}]$ and $[\varphi_{\mbox{\tiny ship}}]$ can be glued in a variety of ways. Definition \ref{def:reg_ins} is \textit{essentially} relational: any variety will be solely a variety of relations between the subsystems. In the ship case, this variety is classified by Galilean transformations, i.e. $I$ has a 1-1 correspondence with the Galilean group, as we saw in \eqref{eq:seaship}.  This example illustrates how a transformation taking $ [\varphi]^{(i')}\rightarrow  [\varphi]^{(i')}$, i.e. altering the physical ($\G$-invariant) universal state, could recover DES as defined by Definition \ref{def:DES}.

  In the following, we will see that there are circumstances in which  Yang-Mills subsystems indeed admit transformations with DES due to GNSS in this full sense.  Namely,  for 
certain regional gauge-invariant data which can be glued together i.e. composed to give a physically possible universal state, there 
remains a \textit{residual variety} of universal gauge-invariant data obtained from 
this gluing. This variety is parametrized by regional \textit{rigid} 
transformations, not regional malleable transformations, and is encoded by the external action of a finite-dimensional Lie group on a subsystem.

In other words, in some cases there 
is a `regional rigid symmetries'-worth of universal states which are \textit{regionally} 
gauge-equivalent to each other but have a  \textit{relational} physical distinctness, i.e. are physically distinct (not universally gauge-equivalent) due to relational differences. The particular structure of $I$ will be that of a rigid subgroup of the set of gauge transformations, but it will depend on the regional field content in each case. It is the field content which identifies  the rigid subgroups of the full infinite-dimensional malleable group that have DES.

  In Yang-Mills theory,  GNSS will occur only in conjunction with those conditions which are necessary for the existence of conserved global charges, as related to the rigid subgroup $I$.\footnote{Namely, it will occur  only for reducible configurations---those which have stabilizing gauge transformation (analogous to non-trivial Killing fields for a spacetime metric)---in which case $I$ is the group of reducibility parameters (analogous to the isometry group of a spacetime metric) \cite{BarnichBrandt2003}. \label{ftnt:reducible}} This procedure thus establishes a link between what is known as an \text{indirect} consequence of gauge---the conservation of charges---and a `direct' one (DES). 


\section{Finding GNSS}
    In this section I will explore the definition of  DES as under-determination of universal \textit{gauge-invariant} data from regional \textit{gauge-invariant} data.  That is,  I will explore DES according to GNSS as described in Definition \ref{def:reg_ins}.  I will illustrate this for  Abelian gauge theories (i.e. electromagnetism with a scalar field).\footnote{ I will only discuss the non-Abelian case in the Appendix.  In that case,  a concession must be made: due to the non-linear character of the theory,  gluing takes place at the perturbative level, and so we must specify which underlying  configuration is to suffer the perturbations. (So the index set $I$ would have to be written as $I([\varphi^{\pm}])$ as envisaged just after equation \eqref{eq:2.3}). Nonetheless, the formalism transforms covariantly with respect to gauge transformations of the perturbed configuration and one is able to retain, for the non-Abelian context, all the interesting results obtained in this Section. This sounds  like the BRST treatment of gauge theories (see \cite{HenneauxTeitelboim} for a review), by which one retains global transformations even if one eliminates the degeneracy in the propagator of the theory through a perturbative gauge-fixing. And indeed, the tools used in  this work in the non-Abelian context recover the properties of BRST ghosts; see \cite{GomesRiello2016} and  \cite[Sec 3.1]{GomesHopfRiello} for more on this recovery. }  For this simple case, I will explicitly show that the whole state is underdetermined by the regional states, and that the ensuing variety of universal states is equivalent to a copy  of (i.e. is parametrized by) the charge group ($\mathrm{U}(1)$).  I will thus prove  my main claim that there is a regional (or subsystem) rigid group of symmetries with (relational) DES, emerging from residual variety, as per Definition \ref{def:reg_ins}.
 

I will organize this section  as follows. 
 Section \ref{sec:notation}  will introduce the necessary notation.  Section  \ref{sec:setup} sets up the remaining tools for the procedure, giving an introduction to our use of gauge-fixing and gluing.  
   In Section   \ref{sec:gluing_regionals} I describe precisely how `external sophistication' is deployed to facilitate the  gluing of the  regional physical states. In that section, I vindicate my main claims on the connection between supervenience and relational DES: viz, that there is a rigid group of regional symmetries parametrizing by the residual variety of \textit{universal} physical states which are composed by identical \textit{regional} physical states.

   \subsection{General Notation }\label{sec:notation}

  We  are given a manifold $\Sigma$, which for our illustrative purposes in this Subsection will represent a space(time) endowed with an Euclidean metric.\footnote{The Euclidean metric provides a simpler interpretation of the results on gluing, in the next section. See \cite{GomesRiello_new} for more on this topic, and for how one can import the relevant results to the Lorentzian signature, ``3+1'' context. In any case, the philosophy of physics literature on DES ignores the contrast between Euclidean and Lorentzian signatures.  }   I will also assume $\Sigma$ is closed, that is, compact and without boundary. Given a charge group $G$, i.e. the finite-dimensional Lie-group characterizing the theory, the group of gauge transformations  is $\G=C^\infty(\Sigma, G)$. 
 The gauge field $A$ and its gauge-transformed  $A^g$ are given by Lie-algebra valued space(time) 1-forms.  
 
 In the main text of the paper, I will only consider the simpler case of Maxwell electrodynamics coupled to scalar Klein-Gordon theory. Thus, for the Abelian case, the structure, or charge, group is $G=U(1)$ and we  write 
 \be\label{eq:EM_trans}A^g=A+i\mathrm{grad}(g)\quad \text{and} \quad \psi^g=\exp(ig)\psi.\ee
 Here $A\in C^\infty(T^*\Sigma, \bb R )$ (a smooth Lie-algebra valued 1-form on $\Sigma$) and  $g\in C^\infty(\Sigma, \bb R)=C^\infty(\Sigma, \mathrm{Lie}(G))=:\fG$, an infinitesimal gauge transformation. This means I am taking a slight short-cut and working with the Lie-algebra $\mathfrak{u}(1)\simeq \bb R$ as opposed to the group $U(1)$.  But this distinction is unimportant for our context ---the Abelian case---and purposes:---finding GNSS.\footnote{In the Abelian case we will be mostly concerned with, the relation between the Lie algebra element $g\in \mathfrak{u}(1)$ and the group element, $\xi\in G$, is:  $g=-i\ln \xi$, or $\xi=\exp(ig)$. This translation can be applied at any point in the following computations. Using the Lie-algebra rather than group is useful in translating our results to the non-Abelian case, since there we cannot work directly with the group. }  The non-Abelian case is treated in the Appendix.
 
I will also assume that the manifold $\Sigma$ is endowed with a Riemannian metric, $g_{ij}$, and that 
 it is decomposed into two regions (cf. Figure 1): $\Sigma_\pm$, with boundary $\pp \Sigma_\pm=:S= \Sigma_+\cap\Sigma_- $.
 \begin{figure}[t]\label{fig:regions}
		\begin{center}
			\includegraphics[width=.3\textwidth]{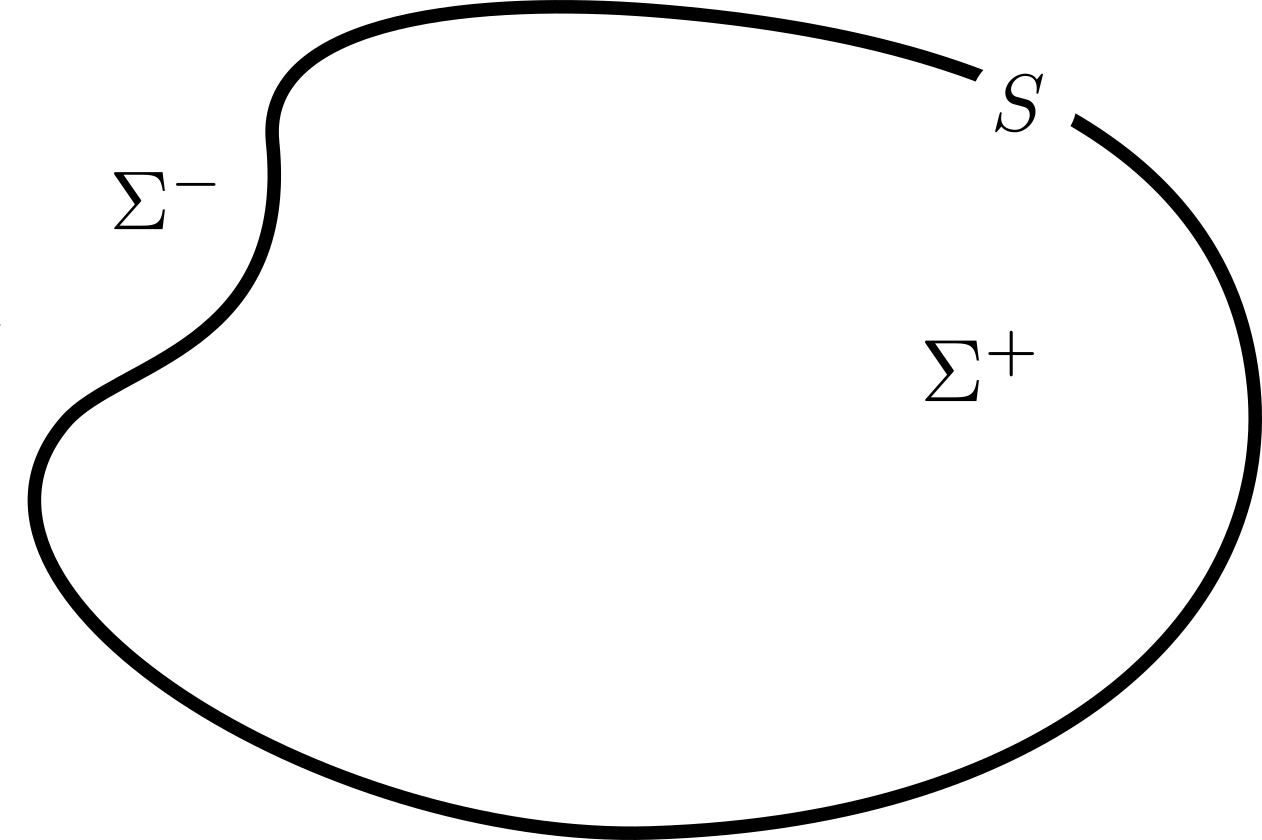}
			\caption{The two spacetime regions $\Sigma^\pm$, separated by a manifold with codimension one, $S$.}
		\end{center}
	\end{figure}
The $\Sigma_-$ piece is what is usually labeled `the environment', but here will play the role of another subsystem (as for relational DES; cf. footnote \ref{ftnt:non_rel_DES}). 	For now,  each of  $\Sigma, \Sigma_\pm$ is assumed topologically trivial; and, if any of these manifolds is not compact, then all the fields on  them will be restricted to have some suitably fast fall-off rate.\footnote{In Figure 1, one could  think of $\Sigma_-$ as a collar around $\Sigma_+$; this would not block our treatment.  The only complication would be to then consider further boundary conditions on $\Sigma_-$, and so on. We therefore restrict our attention to the case where $\Sigma_-$ encompasses the ``rest of the universe'', i.e. the entire  ``environment'' in the language of GW.\label{ftnt:reference_sigma}}

	  I will denote the regional, unquotiented configuration spaces of each field sector (gauge field $A$, matter field $\psi$,  and doublet (gauge and matter fields) $\varphi$) as  $\mathcal{A}_\pm$, $\Psi_\pm$ and $\F_\pm$ (respectively).    I will omit the subscript $\pm$ for the corresponding universal configuration spaces (i.e. $\mathcal{A}, \Psi, \F$, respectively). The restricted groups of gauge transformations  (i.e. smooth maps from regions of the manifold into $G$)  will be denoted in analogous fashion: $\G_\pm=C^\infty(\Sigma_\pm, G)$, and all abstract quotient spaces are denoted by  square brackets, as in $[\Phi^\pm]:=\Phi^\pm/\G_\pm$,  and $[\F]:=\F/\G$. 
	
	If  the fields compose smoothly,  the left hand sides in the following equation are both smooth fields:
\be\label{eq:heavi_dec} A=A_+\Theta_++A_-\Theta_-, \qquad\text{and}\qquad \psi=\psi_+\Theta_++\psi_-\Theta_-.
\ee
Here $\Theta_+$ and $\Theta_-$ are the characteristic functions of the regions $\Sigma_+$ and $\Sigma_-$ i.e. they are distributions: unity in the region, and zero outside, with some conventional value at the boundary which is immaterial for our purposes. Smoothness requires equality of the following quantities at $S$:   $(\pp^{{n}} A_+){}_{|S}=(\pp^{{n}} A_-){}_{|S}$ and mutatis mutandis for $\psi$ in place of $A$; where the superscript ${n}$ denotes all derivatives: first order, second order, i.e. ${n}=1,2$, etc.; and for any combination of independent directions, i.e. $x,y, z$ parametrizing $\Sigma$. We will understand all equalities at $S$, written for simplicity like $A_+{}_{|S}=A_-{}_{|S}$, to be such an equality.

  \subsection{DES as  GNSS: set-up}\label{sec:setup}
   At first sight, the question we face in this Section is much simpler than the previous literature's construals of DES: how do we make sense of the equation (cf. \eqref{eq:reg_ins}): 
  \be\label{eq:gluing_quotients}    [\varphi]_{(i)}=[\varphi^+]\cup^S_i [\varphi^-],\qquad i\in I \quad  ?\ee
But  there are two related obstacles.
  
   First, the elements of $[\mathcal{A}]$, being orbits of the gauge field space $\cal A$ under the action of $\cal G$, are not directly parametrizable. Second,  the only criterion for gluing quotients  employs  representatives, i.e. elements of $\A$. That is, there is no composition of physical states that is not formulated in terms of the composition of representatives.
   
    The first step in overcoming these issues is to ``gauge-fix'': that is,  to select a class of elements of $\mathcal{A}$ which uniquely represent elements of $[\mathcal{A}]$ (cf.  \cite[Sec. 2]{GomesStudies}). A selective class fixes further conditions which the representatives must satisfy.  The second step will consist in  exploiting external sophistication, as advocated in Section \ref{sec:soph}.

  Given  $[A_\pm]\in [\mathcal{A}_\pm]$,  for any two  representatives  $A_\pm\in \A$ of $[A_\pm]$, the condition of composition can then be translated into the following gluing condition: there exist  gauge transformations, $g_\pm\in \G_\pm$, such that the gauge-transformed representatives smoothly join (cf. \eqref{eq:EM_trans}):
\be\label{eq:bdary_cont} A_+^{g_+}{}_{|S}=A_-^{g_-}{}_{|S}\Rightarrow (A_+-A_-)_{|S}=\mathrm{grad}( g_+-g_-)_{|S}.
\ee
 If there are also matter fields that are non-zero on $S$, the corresponding relation to \eqref{eq:bdary_cont} is:
\be\label{eq:bdary_cont_psi} \psi_+^{g_+}{}_{|S}=\psi_-^{g_-}{}_{|S}\Rightarrow \exp(i g_+)\psi_+{}_{|S}=\exp(i g_-)\psi_-{}_{|S}.
\ee


 To ensure that we retain the full physical content of the regional states and the capacity to relate arbitrary configurations, there must be no \textit{prior} restrictions on the gauge transformations at the boundary.\footnote{While such a truncation is standard in  the literature concerning gauge theory in asymptotic regions (cf.  \cite{ ReggeTeitelboim1974, Balachandran:1994up, Giulini_asymptotic, strominger2018lectures}), in that context there are no subsystems that should be glued back together, and the environment is \textit{not} on a par with the subsystem. See  \cite{DES_gf} for a treatment of finite regions with such an assumption, there called `the externalist notion of boundary'  and \cite{RielloSoft} for a recovery of the asymptotic results, including the so-called soft charges, using the present framework. \label{ftnt:truncation}}  This is crucial, both conceptually---why should the ``redundant descriptive fluff'' at the boundary be any different  than in the bulk?---as well as technically.  Our focus on gauge-invariant quantities  
 thus allows a pure gauge discontinuity at the boundary.

  In other words, from the perspective of the subsystem-instrinsic information---as in condition $\mathit{2}$ of Definition \ref{def:DES}---the only criteria for the composition of regional representatives is   whether under some appropriate regional gauge transformations  the resulting  doublets of  regional representatives smoothly compose.

 \subsection{Reduction, sophistication, and gluing}\label{sec:gluing_regionals}
This is the most important technical Section of the paper, in which I illustrate the conceptual tools developed so far. In Section  \ref{sec:reduction}, I will show how one reduces, or fixes a selective class of representatives for the gauge fields, through a gauge-fixing.\footnote{In the Abelian case, the covariance property under transformations of the perturbed configuration need not be flagged explicitly, cf. footnote \ref{ftnt:abelian} below.} In Section  \ref{sec:c} I will describe how external sophistication is important for gluing. And finally, in Section  \ref{subsec:gluing} I describe the residual physical variety. 

\subsubsection{The projection $h$ introduced}\label{sec:reduction}
Given the regions $\Sigma_\pm$, we will consider two  states,  $\varphi^\pm=(A^\pm, \psi^\pm)$.  We need to  uniquely characterize the physical content of such states.  

Let us first focus on the gauge-fields, $A^\pm$. We will return to the matter fields in \ref{subsec:gluing}.
Thus we must fix unique representatives of $A^\pm$, through a projection:\footnote{Here $h$ stands for \textit{horizontal}. Although we will not need to introduce the entire field-space principal fiber bundle formalism here \cite{GomesRiello_new}, this is where `horizontal' comes from: horizontal directions are essentially a choice of non-gauge directions in field space transforming covariantly along the fiber. The word `horizontal' is more appropriate in the non-Abelian case: in the Abelian, horizontal directions are integrable, and correspond rather to a (covariant) foliation of the field-space $\cal A$, i.e. to a $\G$-covariant family of gauge-fixings. \label{ftnt:abelian}}
 \begin{align}
 h^\pm:\A_\pm&\rightarrow \A_\pm\nonumber\\
  A_\pm&\mapsto h^\pm[A_\pm]=: h^\pm_A
\label{eq:h_pm_bdary}  \end{align}
   where $h^\pm_A$ will uniquely represent (i.e. is in the image of all elements of) the equivalence class $[A_\pm]$. 
   
   But we must maintain our ability to describe the possible gluing of the $h^\pm_A$ through an analogue of \eqref{eq:bdary_cont}. Thus $h^\pm_A$ must be gauge-invariant intrinsically---reflecting internal reduction---while still allowing gauge transformations to act, ``extrinsically'', on them---reflecting external sophistication. That is, in the following,  I endorse `reduction' (cf. Section \ref{sec:soph}) for determining the regional and universal physical content. But, to describe gluing, I endorse `external sophistication': i.e. allowing all the different representations of the same regional physical content to be counted as isomorphic but \textit{not} identified from the outside view. 
   
Technically, these demands imply we should look for a projection  $h:\mathcal{A}\rightarrow\mathcal{A}$, as opposed to a reduction, $\mathsf{red}:\mathcal{A}\rightarrow [\mathcal{A}]$. In \cite{GomesStudies, 
DES_gf} the construal of a gauge-fixing as a projection $h$, and not as a quotient, was argued to be fundamental for the gluing of regions: for both $h$ and $\mathsf{red}$ are required to be gauge-invariant with respect to gauge-transformations on the common domain, $\mathcal{A}$, i.e. $\mathsf{red}(A^g)=\mathsf{red}(A)$ as well as $h(A^g)=h(A)$,  but only the projection $h$ allows further transformations to be enacted on its range.

 First, \textit{without loss of generality}, we can find linear projections $h^\pm:\A_\pm\rightarrow \A_\pm $ such that their images satisfy:
 \be\begin{cases}
  \mathrm{div}(h^\pm_A) &=0 \\
   s\cdot h^\pm_A& =0\end{cases} \label{eq:hor_conds}\ee
where $s$ is the normal to the boundary $S$, and $\cdot$ is induced by the inner product on $\Sigma$. It might seem surprising that we can restrict our attention to such constrained boundary conditions on $ A_\pm$ and yet still encompass the entire gamut of possible regional physical states. The reason for this is that any other state---including ones with boundary behavior different from \eqref{eq:h_pm_bdary}---differs from such a $h^\pm_A$ by a unique regional  gauge transformation; again, a regional gauge transformation that is possibly non-trivial at the boundary.

There are many ways of gluing (as we will see in Section \ref{sec:c} below), which will give rise to many possible universal representatives of the state. But GNSS refers to the univesal \textit{physical} state as well as to the regional physical states. Thus, after gluing, we need to resolve the mapping between regional and  universal physical states, and so we need to determine physical universal states from glued regional physical states. That is, we need to eliminate the plethora of possible extrinsic gauge transformations as well as the intrinsic ones.

To discern or distinguish the possible universal physical states, we apply the corresponding gauge fixing for  closed---compact without boundary---regions. In the absence of boundaries, i.e. for the universal state $A$, instead of \eqref{eq:h_pm_bdary} we have: 
\begin{align}
 h:\A&\rightarrow \A\nonumber\\
  A&\mapsto h[A]=: h_A
\label{eq:h}  \end{align}
and the significance of \eqref{eq:hor_conds} reduces to the familiar statement that the condition  
\be\label{eq:Landau}
\mathrm{div}( h_A) =0;\ee
 is a bona-fide (partial)  gauge condition, called the (Euclidean) Lorentz gauge, or, in the non-Abelian setting,  the (perturbative) covariant Landau gauge. It is `partial' because a different choice, related to $h$ by a constant shift, would still satisfy \eqref{eq:Landau}. It implies that the representative of the universal physical state is only determined up to a global phase shift. This  will be important in what is to come.
 
 The condition \eqref{eq:Landau} will fix the extrinsic gauge transformations, so that we find the sought-for correspondence between regional physical states and universal physical states. 

Let us add some detail to this procedure. 

\subsubsection{Internal reduction and external sophistication: option (c)  realized.}\label{sec:c}
We know that each $h$ and $h^\pm$ yields a unique element in each orbit because of the way  the projection $h:\A\rightarrow \A$ works by exploiting gauge transformations. For instance, in the global case, for $\U(1)$:
 \be\label{eq:h[A]}h[A]:=A+i\,\mathrm{grad}(i\nabla^{-2}(\mathrm{div}(A)))=A^{\sigma[A]},\ee
 where the functional
 \be\label{eq:sigma}  \sigma[A]:=i\nabla^{-2}(\mathrm{div}(A))
 \ee
 is the unique solution of the equation: 
\be\label{eq:h_div} \mathrm{div}(h[A])= \mathrm{div}(A^{\sigma[A]})=0.
\ee 
Moreover, it is easy to see from \eqref{eq:h[A]} that the $h[A]$ satisfies $h[A^g]=h[A]$, $\forall g\in \G$. Thus $h$ is a complete, gauge-invariant functional,  uniquely representing each equivalence class; we have one, and only one $h^A:=h[A]$ per orbit of the gauge group.

And a similar $\sigma^\pm$, with analogous properties,  exists for  for $h^\pm$ obeying  the regional equations \eqref{eq:hor_conds}. 

Therefore we can now distinguish two sorts of action of $\G$: a subsystem-intrinsic action and a subsystem-extrinsic one.  Subsystem-intrinsic transformations will map between the members of the same equivalence class, whereas the extrinsic ones act as transformations between the representatives of these equivalence classes.\footnote{ It is instructive to compare the two possibilities of action of $\G$  to the use of homotopy type theory (HoTT) in gauge theory,  as advocated by \cite{Ladyman_DES}. Ladyman says HoTT ``both (a) distinguishes states conceived of differently even if they
are subsequently identified, and (b) distinguishes the identity map from non-trivial transformations that nonetheless might be regarded as delivering an
identical state''. Here we have two sorts of transformations: the subsystem-intrinsic  one, $A\mapsto A^g$, which does not change $h[A]$---satisfying Ladyman's (b)---, and the subsystem-extrinsic one, that does the work of Ladyman's (a). }\\

In more detail, we have:
\paragraph{Subsystem-intrinsic gauge transformations}
Given  $h^\pm:\mathcal{A}_\pm\rightarrow\mathcal{A}_\pm$, where domain and range are seen as distinct, but isomorphic, spaces, a subsystem-intrinsic gauge-transformation is just a gauge-transformation acting on the domain of $h$. It maps between members of the same equivalence class.\\

The label `intrinsic' stands in opposition to `extrinsic'. Intrinsic gauge transformations are all that is needed for a unique description of the entire Universe, since there is nothing external to the entire Universe.  

But if we have more than one subsystem and we want to satisfy the gluing condition \eqref{eq:bdary_cont}, we may need to change the representative of the equivalence class $[A_\pm]$---from the outside, as it were.

\paragraph{Subsystem-extrinsic gauge transformations}

 We can define \textit{subsystem-extrinsic} gauge transformations $g^{\text{ext}}$, as those transformations which act on the \textit{range} of $h$ as 
\be\label{eq:external_gt}
h^\pm_A\mapsto h^\pm_A+i\,\mathrm{grad}(g_\pm^{\text{ext}}).\ee 
Of course such a transformed field would no longer satisfy \eqref{eq:hor_conds}.

That is, subsystem-intrinsic gauge transformations are defined as those acting on  the field configurations in the domain of the projection, whereas the subsystem extrinsic act on its range. Once we have eliminated redundancy and fixed a 1-1 correspondence with $[A_\pm]$, the image of $h^\pm$, i.e. $h^\pm[\mathcal{A}_\pm]\subset\mathcal{A}_\pm$, is invariant with respect to gauge transformations acting on its domain, but we can still change representatives by acting on its range, $\mathcal{A}_\pm$.

\paragraph{Gluing}

We are given the physical content of the regional configurations as (in terms of) their projected representatives $h^{\pm}_A$, and while these representatives $h^{\pm}_A$ might not smoothly join,  they may still  jointly  correspond to a physically possible universal state. 
The existence of subsystem-extrinsic gauge transformations smoothening out the transition between $h_A^+$ and $h_A^-$ is a necessary and sufficient condition for their compatibility.

 That is, the $h_A^\pm$ determine whether they can be smoothly joined by subsystem-extrinsic  gauge transformations. Following \eqref{eq:bdary_cont}, the condition is that subsystem-extrinsic gauge transformations $g^{\text{ext}}_\pm$ exist such that (in spacetime index-free notation\footnote{Using indices, the equation is: $(h^\mu_+-h^\mu_-)_{|S}=i\partial^\mu( g^{\text{ext}}_+- g^{\text{ext}}_-)_{|S}$. }):
\be 
(h_A^+-h_A^-)_{|S}=i\mathrm{grad}( g^{\text{ext}}_+-g^{\text{ext}}_-)_{|S};\label{eq:bdary_cont2}
\ee  which is the appropriate rewriting of the gluing condition \eqref{eq:bdary_cont}.

However, as mentioned, this is not enough to establish a correspondece between universal and (the doublet of) regional physical states: although we have the physical representatives on the regional side, we do not yet have them on the universal side of the correspondence. 

\subsubsection{Establishing the main claims}\label{subsec:gluing}
 
 Finally, we must know which of these gluings give rise to physically distinct universal configurations. Thus, we need  to eliminate the redundancy of subsystem-extrinsic gauge transformations and get a unique representative of the universal state. We first write: 
 \be\label{3} h_A:=(h_A^++i\mathrm{grad}(g^{\text{ext}}_+))\Theta_+ +(h_A^-+i\mathrm{grad}( g^{\text{ext}}_-))\Theta_-
 \ee
 where $g^{\text{ext}}_\pm$ obey \eqref{eq:bdary_cont2}, and then we apply the reduction of the universal state, through \eqref{eq:Landau}, i.e. demanding that $\mathrm{div}(h_A)=0$ (we assume the universe has no boundary). Imposition of this divergenceless condition on \eqref{3} almost uniquely fixes the solution: $g^{\text{ext}}_\pm[(h_A^+- h_A^-)_{|S}]$.\footnote{ In both Abelian and non-Abelian case, $g^{\text{ext}}_\pm$ depends on $h_A^\pm$ only through their difference at the boundary: $( h_A^+-h_A^- )_{S}$ (in the non-Abelian case, each $g^{\text{ext}}$ depends also on  its respective regional gauge field, e.g. $g^{\text{ext}}_+[A_+, (h_A^+- h_A^-)_{|S}]$). For illustration purposes, I display the solution here: 
$$\ln g^{\text{ext}}_\pm = \zeta_{(\pm)}^{ \pm\Pi}\quad\text{with}\quad \Pi=\Big(\mathcal{R}^{-1}_+  + \mathcal{R}^{-1}_-\Big)^{-1}\left(  (\nabla^2_S)^{-1}\mathrm{div}_S( h_A^+-h_A^- )_{S}\right), $$
where the subscript $S$ denotes operators and quantities intrinsic (i.e. pulled-back)  to the boundary surface $S$ (and since normal components of $h_A^\pm$ match, $( h_A^+-h_A^- )_{|S}=( h_A^+-h_A^- )_{S}$); $\zeta_{(\pm)}^u$ is a harmonic function on (respectively) $\Sigma^\pm$ with Neumann boundary condition $ \pp_n \zeta^u_{(\pm)}=u$, and $\mathcal{R}$ is the Dirichlet-to-Neumann operator. For the meaning of these operators, and also the analogous solution for the general non-Abelian Yang-Mills gauge theories, see \cite[Sec. 4]{GomesRiello_new}.}  In the $\U(1)$ case, there is an under-determination by a constant  extrinsic gauge transformation in each region; let us call this indeterminacy $(c^+, c^-)$. Thus $g^{\text{ext}}_\pm[h_A^+, h_A^-]$ is fixed \textit{up to} the addition of constants, $c_\pm\in \bb R$. We note that as expressed,  this is an indeterminacy at the level of the Lie-algebra; for the Lie-group, we would get $\exp(i c_\pm)\in [0,2\pi]$.

To get to the main claims of the paper, we must now include charged matter fields in our description. 

\paragraph{Charged matter fields}
 
 First, note that the discussion so far  focused on the gauge fields, $A$. But including matter fields is not difficult: since matter fields `co-rotate' with $A_\pm$, they just get ``taken for the ride'' by the fixing  of $\sigma[A]$. That is,  since the $\sigma_\pm$  are unique, a given  doublet, $ \varphi_\pm=(A_\pm, \psi_\pm)$ is also projected to a unique representative:
 \be\label{eq:dressed_fields} (h^\pm_A, h_\psi^\pm)=(A_\pm+i\mathrm{grad}(\sigma_\pm[A_\pm]), \exp{(i\sigma_\pm[A_\pm]}\psi_\pm),
 \ee
where $\sigma_\pm$ is the regional version of \eqref{eq:sigma}.\footnote{ In the asymptotic flat case, the universal functional $\exp{(i\sigma[A])}\psi$ is known as the ``Dirac dressed electron'' \cite{Dirac:1955uv}. It is an electron that is ``dressed'' by an appropriate Coulombic tail, rendering the electron also gauge-invariant (as can be easily checked from the gauge-covariant transformation properties of $\sigma$). The function $h_\pm$ is known as the ``radiative'' projection of the photon. Here we have extended both notions to finite bounded regions (see \cite{GomesHopfRiello, GomesRiello_new}).}

Since we  find \textit{unique} representatives for the full regional physical field content,  we can identify (using the notation`$\equiv$') the content with the representative:
\be\label{eq:unique_regs}
[\varphi_\pm]\equiv (h_A^\pm, h_\psi^\pm),
\ee
and take these as the starting point for gluing.  

  Inputting \eqref{eq:unique_regs} into the universal state \eqref{3}  and solving for \eqref{eq:Landau}, we obtain $g^{\text{ext}}_\pm[h_+, h_-]+c_\pm$. Thus, we have: 
\be\label{eq:final_2} [\varphi]_{(c^+, c^-)}\equiv \Big((h_A^++i\mathrm{grad}( g^{\text{ext}}_+)), \exp{i(g^{\text{ext}}_++c_+)}h_\psi^+\Big)\Theta_+  +(+\leftrightarrow-)
\ee
 where the term $(+\leftrightarrow-)$ is identical to the first, mutatis mutandis for $+\leftrightarrow-$, and I have omitted the dependence of $g^{\text{ext}}_\pm$ on $h^\pm_A$. The important point to understand is that, although the degenaracy in gluing, $c_\pm$,  has no effect on the gauge field, it will add a phase, or rotate, the matter part. \\
 
 But now we must consider two cases: either the $\psi_\pm$ vanish at $S$, or they don't. 
 
 Assume first that $\psi_\pm$ vanish at $S$. We then  have no further constraints and thereby obtain a 2-parameter family of universal  states, \eqref{eq:final_2}, parametrized by $\exp(ic_\pm)$. 
In other words, given regional physical states $[\varphi_\pm]$, here a conjunction of e.g. a transverse projection of the photons and a Dirac dressing of the charges, we can glue them to form a two-parameter collection of universal states.

Naively, this would give us \textit{two} copies of $\mathrm{U}(1)$,  parametrizing the  universal physical states that are compatible with the regional physical contents. But of course, if $c_+\equiv c_-$, we have a global constant phase shift (which precisely matches the expected left-over under-determination of the universal physical state $[\varphi]$ by \eqref{eq:Landau}) which does not change the universal physical state. Therefore, we are left with a residual physical variety parametrized by the difference, $\bar c:=c_+-c_-$,  which is insensitive to any global phase shift. To sum up, if the matter fields vanish at the boundary, we obtain a residual variety that is isomorphic to a single copy of $\mathrm{U}(1)$ and in fact can be identified with relative, regional, rigid phase shifts.

Assume now  that $\psi_\pm$ do not vanish at $S$. Then not every boundary value is allowed: the two sides must differ by a phase at the boundary, and this phase difference needs to match the gauge transformation required to glue the gauge potentials. That is, by \eqref{eq:bdary_cont_psi},
 \be\label{4}\exp{i((g^{\text{ext}}_++\sigma_+)-(g^{\text{ext}}_-+\sigma_-)+(c_+-c_-))}\psi_+{}_{|S}=\psi_-{}_{|S}
 \ee
 where $(g^{\text{ext}}_++\sigma_+)-(-\leftrightarrow+)$ is a fixed functional of the gauge fields, and $c_\pm$, which were entirely under-determined previously by the gauge fields $A$, will now completely fix the quantity $(c_+-c_-)$ by  \eqref{4}. Thus, if the matter fields do not vanish at the boundary (and are compatible with each other and with the gauge fields, satisfying \eqref{4}), there is no variety left, since the compatibility equation \eqref{4} completely fixes the difference $\bar c$. \\

Finally, as a corollary of these constructions,  we are able to state our main result of this Section for gluing physical states: 
\begin{theo}[Rigid variety for $\mathrm{U}(1)$]\label{cor}
  For electromagnetism as coupled to a Klein-Gordon scalar field in a simply-connected universe: given the  physical content of two regions, $[\varphi_\pm]$, for matter vanishing at the boundary but not in the bulk of the regions, the universal state is underdetermined, resulting in a residual variety parametrized by an element of $\mathrm{U}(1)$. In the notation of \eqref{eq:reg_ins} and Definition \ref{def:reg_ins}: 
\be\label{eq:final_claim}
[\varphi_{(i)}]=[\varphi_+]\cup_{(i)}[\varphi_-],\quad i\in I\simeq\mathrm{U}(1)
\ee
where the particular action of $\mathrm{U}(1)$ is that which leaves the gauge-fields invariant, but not the matter fields.\footnote{In this phrasing, the variety is more widely applicable, e.g. to non-Abelian fields. For the Abelian case, it just means we take the group of constant phase shifts.} 
\end{theo}

Thus we have found, in the case of  electromagnetism  coupled to a Klein-Gordon scalar field, a \textit{rigid variety} through GNSS, as per Definition \ref{def:reg_ins} in Section \ref{sec:intro_variety}. Moreover, by fixing a reference subsystem ($\Sigma_-$), we can construe this $\mathrm{U}(1)$ as acting as a group of (rigid), regional symmetries over $\Sigma_+$, that shift the phase of matter but do not affect the gauge potentials (as in the `t Hooft beam-splitter,\footnote{: a beam-splitter with two arms (containing electron fields)---the equivalent of our regions---and the  phases are individually shifted by a constant in each  arm. This phase difference can alter the interference pattern, and thus, it serves as an example of a subsystem gauge transformation with DES. \label{ftnt:thooft}} \cite{thooft}). That is, we have constant phase shifts.

In direct analogy with Galileo's ship---where we also recognize  GNSS as encoded by the (external) action of a finite-dimensional group---we have fully vindicated our main claims: namely, that GNSS is a source of empirically significant subsystem symmetries.

Agreed, the non-Abelian Yang-Mills case is more complicated: non-linearities render the corresponding $I$ of the equation corresponding to \eqref{eq:final_claim} dependent on the physical state. Nonetheless, our constructions \textit{are} valid at a perturbative level, i.e. one needs to first fix a ground state and then perturb it. If the perturbed state is the `vacuum', i.e. $[ A^*]$ for $ A^*=0$, then we recover the full Lie algebra of the gauge group through the analogue of \eqref{eq:final_claim}. In that case too, the particular rigid subgroup is identified as the one that leaves the perturbations invariant but not the matter sources.

\section{Comparison with GW and BB}\label{sec:GW_BB}
 Now that our work is done, I will draw several lines of comparison to the approaches of \cite{ BradingBrown,GreavesWallace}. Their positions were introduced briefly in Section  \ref{sec:DES_debate} in sparse detail, but we will not need more than that. In Section  \ref{sec:BB} I briefly gloss BB's argument against DES; it relies on an a priori condition on the possible doublet of regional gauge transformations: that they smoothly join.  In Section  \ref{sec:GW}, I briefly gloss GW's arguments for DES.    In Section  \ref{sec:rigid}, I will show, contra GW, that the lack of DES for general malleable symmetries still allows for the context-dependent
identification of some of their rigid subgroups---which do have DES.
 
\subsection{Brading and Brown: regional gauge transformations must match}\label{sec:BB}

As the quotation at the end of Section  \ref{sec:DES_debate} illustrates,  BB take any two regional gauge transformations which fail to coincide at the boundary to be disallowed.
In discussing the t'Hooft beam splitter (cf. footnote \ref{ftnt:thooft}), they conclude:
\begin{quote}
 ``The only remaining option is to consider a region where the wavefunction can be decomposed into two spatially separated components, and then to apply local gauge transformation to one region (i.e. to the component of wavefunction in that region, along with the electromagnetic potential in the region) and not to the other. But then either the transformation of the electromagnetic potential results in the potential being discontinuous at the boundary between the `two subsystems', in which case the relative phase relations of the two components are undefined (it is meaningless to ask what the relative phase relations are), or the electromagnetic potential remains continuous, in which case what we have is a special case of a local gauge transformation on the entire system.''  (p. 656)
\end{quote}
BB are right in one respect: once you have a universal configuration one \textit{can not} apply gauge transformations which are discontinuous (e.g. produce a delta function) at the interface.\footnote{See p.79 and 82 \cite{GreavesWallace} for their endorsement of continuity of universal gauge transformations.} On the other hand, it is also true  that one \text{could} have regional configurations being acted upon by regional gauge transformations  which don't match at the boundary.\footnote{See p. 83 of \cite{GreavesWallace} for their endorsement of \textit{this} point.} In this second instance, one aims to take the regional subsystem
intrinsically and have non-matching gauge transformations applied to them before
gluing.

In other words,  in their quotation, BB ignore the fact that the formulation of DES, through property $\mathit{2}$ of Definition \ref{def:DES}, requires only information intrinsic to a subsystem---which is why we related it to supervenience of the universe on its subsystems. All hands agree that, from the viewpoint of the universe  one cannot have a  gauge transformation which is discontinuous at the boundary. From this viewpoint it is true that $g_+=g{}_{|\Sigma_+}$ and $  g_-=g{}_{|\Sigma_-}$; i.e. that  the regional gauge transformations are mere restrictions of a universal gauge transformation.  And they are right: such an assumption would pre-empt any search for DES. 

 But these conclusions are unwarranted. For the topic  of supervenience, one \textit{starts from the regional states} and then composes them. From this perspective, it is the \textit{effect of the regional gauge transformations} that matter.

 GW spot this error, and assert that what should be fundamental is only  the continuity of the glued gauge and matter fields, $A, \psi$, not of phase shifts. In their words (but my notation) (p. 83): 
\begin{quote}
The key to seeing why
this argument fails is noting that what is given, when we are given the pre and
post-transformed states of the universe, is not a function from spacetime
to the gauge group, but merely the effect of whatever transformation is being
performed on the particular pre-transformation (universe) state $(\psi, A)$.
And if this particular $\psi$  happens to vanish on the overlap region [the boundary $S$], then
nothing about the corresponding gauge transformations $g_+, g_-$ can be ‘read
off’ from their effects on the wavefunction in that region [assuming they are constant near $S$]. It is therefore possible
that the universe transformation being performed might correspond to the effect
of (say)  \textit{some constant gauge transformation $g_+$ in $\Sigma_+$, and a different constant
gauge transformation $g_-$  in $\Sigma_-$, so that there is no way of patching $g_+, g_-$
together} to obtain a single smooth function from the whole of spacetime to the
gauge group. [my italics]
\end{quote}
I fully agree with this verdict, as far as it goes. It means one considers the effect of the regional $g_\pm$ on the subsystems from the intrinsic point of view. One does not take $g_\pm$ as the regional  projections of a discontinuous $g$. But I believe GW are not entirely consistent in applying this approach, as we will shortly see.

Nonetheless, GW are one step closer to an analysis based solely on physical (gauge-invariat concepts than are BB: they take the representatives of the subsystem fields to be important, not the gauge transformations themselves. I will now develop their view, and, in particular, will parse the italicized text above .


\subsection{Greaves and Wallace: non-matching gauge transformations encode DES}\label{sec:GW}

To recapitulate: GW claim the relational DES transformations are in 1-1 correspondence with the following quotient of two infinite-dimensional groups:
\be\label{eq:DESGW_final}\G^{\text{\tiny{GW}}}_{\text{\tiny DES}}(\varphi)\simeq \mathcal{G}(s{}_{|\pp})/\mathcal{G}_{\text{Id}},\ee
 where $\mathcal{G}(s{}_{|\pp})$ are the gauge transformations of the region which preserve the state $\varphi$ \textit{at the boundary} of the region, and $\mathcal{G}_{\text{Id}}$ are the gauge transformations of the region which are the identity at the boundary. 
 

Here is the gist of the argument (reflected in the above quote) leading from Definition \ref{def:DES} to \eqref{eq:DESGW_final} (and reflected in the above quote): certain gauge transformations are not the identity at the boundary and yet they may keep particular boundary states invariant. One can use such regional transformations to obtain a different universal  state, but the composition of the regional gauge transformations is not itself smooth and therefore  does not count as a gauge transformation relating  the initial and final universal  configurations. GW's mistake is that they implement no criteria to establish whether the regional and universal states are indeed physically distinct. As I will show below, this omission allows us to find a simple counter-example, in which \eqref{eq:DESGW_final} is non-trivial and yet the transformations constructed above fail to yield physically distinct states, and thus do not satisfy condition $\mathit{1}$ of Definition \ref{def:DES}. 


In more detail, let $(A_\pm, \psi_\pm)$  be two configurations, one in each region $\Sigma_\pm$,  that join smoothly, and such that $\psi_\pm=0$. Therefore we have the initial universal configuration $A=A_+\Theta_+\oplus A_-\Theta_-$.

 As a condition for these representatives to smoothly compose, we  must have: $A_+{}_{|S}=A_-{}_{|S}$ (where equality here includes equality of derivatives at $S$). But  $A_+{}_{|S}$ has a stabilizer: at the Lie algebra level, this is $g_+{}_{|S}=c_+\neq 0$ such that grad$( g_+){}_{|S}=0$ and where $c_+$ is some constant on $S$. Therefore,  the configuration
$$ 
\tilde A:=A_+^{g_+}\Theta_+\oplus A_-\Theta_-$$
 is still smooth, since $g_+$ doesn't change the value of $A_+$ at the boundary. But  
 $$g_+\Theta_++ g_-\Theta_-=g_+\Theta_+,$$ with $g_-=0$,\footnote{ That is, GW take $g_-\equiv 0$:  they take the environment gauge transformations to be the identity, or to serve `as a reference' for the gauge transformations of the subsystem, and the original $g_+$ to be the identity. This is a bit  confusing, since stipulating the value of $g$ does not usually fix a selective class: given $A$, one cannot use this condition to assess whether $A$ belongs to the selective class. Here, since we are only interested in a counter-example to substantiate our criticism of their construal of DES, we will ignore this point (which is not an issue for our formulation through GNSS). } is not a smooth (infinitesimal) gauge transformation (because $g_+$ doesn't vanish at $S$).  GW would conclude from this that $\tilde A$ is not a gauge-transformation of $A$, and that, therefore, $\tilde A$ and $A$ are physically distinct. Moreover, they observe, such a $g_+$ corresponds to an element of \eqref{eq:DESGW_final} and  essentially, they claim,  the same construction would apply for any other element of this quotient group. 

But without dealing with gauge-invariant quantities, GW have no warrant to conclude that $\tilde A$ and $A$ are physically distinct, and therefore no warrant to conclude that condition $\mathit{1}$ of Definition \ref{def:DES} is satisfied. In fact,  it is only true that $\tilde A$ \textit{as it is written}, doesn't appear to be a gauge transformation of $A$. But we can \textit{explicitly} construct a gauge transformation relating $\tilde A$ and $A$ as follows: Let $\tilde g_+:=g_+-c_+$. Now $A_+^{\tilde g_+}=A_+^{g_+}$ (because the constant part has a trivial action on $A_+$), and so:
 $$\tilde A=A_+^{\tilde g_+}\Theta_++ A_- \Theta_-.$$
 But now $g:=\tilde g_+\Theta_++g_-\Theta_-=\tilde g_+\Theta_+$ \textit{is} a smooth gauge transformation (since $\tilde g_+$ vanishes at $S$).  And therefore $\tilde A=A^g$. So $[\tilde A]=[A]$: that is, the physical, gauge-invariant, states are identical and the transformation cannot have empirical significance (condition $\mathit{1}$ of Definition \ref{def:DES} fails). \footnote{ 
 If there was matter in the bulk, our counter-example would fail, because $\psi_+^{\tilde g_+}\neq \psi_+^{g_+}$. But the source of such DES would be just what I described here through Theorem \ref{cor}: namely, as required from the Theorem, there is a stabilizer of the gauge potential but not of the matter field. }

 (Although the above argument was formulated explicitly for electromagnetism,  its extension to the more general cases is straightforward.) 

\subsection{GW and BB's error}\label{sec:GWBB_error}

In my view, neither GW nor BB could have obtained the right characterization of DES, for Definition \ref{def:reg_ins}'s characterization of DES in terms of physical i.e. gauge-invariant, states  was not articulated by either group. Indeed, they both explicitly endorse GSS. On this topic,  GW write:
 \begin{quote}
\textbf{GSS'}: ``Firstly, in doing so we make the assumption that knowing the state of the
subsystem and its environment suffices to specify the state of the total system.
[...] For
example, it is true for Yang-Mills gauge theories in the connection formalism
but not in the holonomy formalism.''  (p. 67)
 \end{quote}
 This assumption, which I have labeled as an alternative statement of GSS,  should not be confused with its converse, which, all hands agree,  holds for all of these systems. Namely, the converse assumption---that  the state of the whole,  uniquely determines the states of the subsystems---is not under dispute since the state of the whole includes all relational information and more. But as we have stressed: given just the intrinsic physical states of the subsystems, it is \emph{not} necessarily the case that there is just one way of putting them together; their conjunction may lack necessary relational information.
 
  The exception GW make for the holonomy formalism is telling here:  how can physical significance depend on the choice of variables?  Indeed, for us it \emph{is} immaterial: for electromagnetism, it is possible, using holonomies, to recover precisely the same results as in Theorem \ref{cor}. For GW, it fails precisely because the holonomy formalism deals with gauge-invariant observables.\footnote{Here I  comment on the relation between GNSS and  Myrvold's `global patchy-non-separability' \cite[p.427]{Myrvold2010}, which they articulated for electromagnetism in holonomy variables (cf. footnote \ref{ftnt:Myrvold1}). In electromagnetism, given the space of loops (smooth embeddings $\gamma:S^1\rightarrow \Sigma$), one can form a basis of gauge-invariant quantities by the holonomies, $\exp{(i\oint_\gamma A)}$ (this can be accomplished more generally for non-Abelian theories using Wilson loops, \cite{Barrett_hol}).  For simply connected regions $\Sigma$ like ours, by composing regional loops  $\gamma_\pm\in\Sigma_\pm$ going in opposite directions at the boundary $S$ it is true, as Myrvold argues, that we recover the gauge-invariant holonomy corresponding to a larger loop $\gamma$ not contained in either region. Therefore any universal holonomy corresponds to a single doublet of regional holonomies. According to Myrvold, separability fails only for non-simply connected manifolds, where the holonomies of $\gamma_\pm$ cannot recover the universal holonomy  of $\gamma$.
  
   I have two comments to make on the relation to the present work: (i) In the absence of matter and in the Abelian setting,  non-trivial topology indeed is the only source of GNSS, and we recover Myrvold's conclusions. But with matter, we can close off  curves which are not loops---which only change by gauge transformations at their ends---by placing charges to cap off the curves,  thereby obtaining gauge-invariant holonomies (cf. \cite[Sec. 4.3.2]{GomesRiello_new}). These sourced holonomies correspond to the residual  variety we found for electromagnetism in Theorem \ref{cor}, which occurs even for simply-connected $\Sigma$ when charged matter is present in the  regions $\Sigma_\pm$.
   
    (ii) Unfortunately, Myrvold's loop composition doesn't work in the same way for the non-Abelian theory: although the appropriate regional loops themselves will compose as curves in the manifold, Wilson loops---giving the gauge-invariant content of the holonomies---involve traces, and the traces ruin the composition properties: the corresponding regional gauge-invariant quantities do not  compose. Our construction (see appendix \ref{app:gluing_YM}) gets around that. 
  
  The conclusions, in Myrvold's nomenclature, are then: global patchy separability fails for non-Abelian theory if and only if: the manifold is non-simply connected, or charged matter is present inside the regions \textit{and} furthermore this charged matter obeys (perturbative) regional conservation laws (i.e. and the perturbed gauge-field has stabilizers).\label{ftnt:Myrvold2}}    

\subsection{How can a rigid subgroup have DES? }\label{sec:rigid}

Now it is our turn to  defend our construction from one of GW's arguments, who judge it impossible that a  `global' (i.e. rigid) gauge transformations, being a subset of the `local' (i.e. malleable), may acquire DES, while none of the malleable do. 

  As expressed in equation \eqref{eq:DESGW_final} for \textit{relational} DES, GW only claim a group isomorphism between $\G^{\text{\tiny{GW}}}_{\text{\tiny DES}}$, whatever it  may be, and a \textit{quotient} group,   ``empirical symmetries correspond 1-1 [...] to elements of a quotient group'' (p. 75).\footnote{The claim  is based on intuitions for gauge theories in the asymptotic regime. But the situation in the asymptotic regime, or for truncated configuration spaces, is more complicated: there, one does not consider environment and subsystem to be on a par as we do for relational DES; the environment truncates the fields and the gauge transformations at the boundary of the subsystem. In \textit{that} case, we \textit{can} obtain a group of symmetries with DES that is isomorphic to the quotient and has  no action on the subsystem states. These results, obtained in \cite[Sec 4.2.2]{DES_gf}, largely recover those presented by GW (erroneously) for relational DES in \eqref{eq:DESGW_final}, and more.\label{ftnt:GW_recover}}
 But  GW do not see the quotient nature of their result as problematic. In fact, they see it as exonerating their notion of relational DES from the charges they make against the orthodox view on DES:  
\begin{quote}
 In any theory that has a malleable symmetry group, the rigid symmetries
remain as a subgroup of that malleable symmetry group. (For example, in
general relativity, the rigid translations and boosts form a subgroup of the
group of all diffeomorphisms.) It is therefore logically impossible that all
rigid symmetries, but no malleable symmetries, can have direct empirical significance; (p. 61)
\end{quote}
(again the quotation is adapted to my nomenclature). That is, their accusation against the orthodox view is that it attributes relational DES only to certain rigid subgroups of malleable groups.  
Recognizing the quotient group as being isomorphic (in 1-1 relation) to the symmetries exhibiting DES, GW argue, assuages this concern.\footnote{``
 The reason lies in a structural difference between the original
problematic claims and our replacements: rather than holding that all elements
of a (malleable) symmetry group have one property while elements of some
subgroup thereof have a contradictory property, we hold that (for any given
subsystem) there is a subgroup of subsystem symmetry transformations (the
‘interior’ ones) that cannot have empirical significance, and that it is elements
of the quotient of the larger group by this subgroup that are \textit{candidates for
correspondence to physical operations}. There is thus no object of which we
assert both that it does, and that it does not, have some given property.'' (p. 87, my italics).
 A rigid subgroup corresponds to physical operations through its action on the configuration variables (e.g. boosts on the ship, or global changes of phase). But how the quotient is to do the same is not specified.  And it is easy to see that if one insists only on the 1-1 relation between the quotient group and the group with DES---i.e. the one which \textit{could} act as  physical operations on a subsystem---there is no unique way of realizing the said physical operations (as there is with Galileo's ship); even if the quotient is finite-dimensional, it will have a continuous infinity of equivalent representations  on the region.  }

But is it  really true  that one cannot endow significance solely to a subgroup of a group? What are \textit{the} rigid translations and boosts of a generic spacetime metric?  Poincar\'e transformations are not well-defined (i.e. geometrically defined) in a generic background metric. They are defined by Killing fields for a Minkowski metric. In that sense, that subgroup is \textit{physically distinguished}, but only  in particular circumstances: it is only meaningful for a Minkowski metric;\footnote{If one is thinking not in terms of active diffeomorphisms, but of coordinate transformations, then indeed, one can single out translations and boosts, but only with respect to that coordinate system.  More broadly, there are generically no constant gauge-transformations: they usually require a global section to be defined. One should also note that although for principal fiber bundles one has a natural action of the charge group, $G$, this is not the case for associated bundles. One cannot define a ``constant''  action of the gauge group: it can only be constant with respect to a given section (see \cite{kobayashivol1}).}
  in specific backgrounds (e.g. Minkowski), there will be a physically well-defined rigid subgroup of the malleable transformations, but this subgroup will be effaced once one moves to generic backgrounds. 
  
Admittedly: if one focuses just on  the group itself, and not on its action on states, one indeed cannot ``pluck out'' a rigid subgroup in any meaningful way.     
  But \textit{contra} GW, it is entirely possible to associate DES only to certain physically meaningful rigid symmetries; the meaning is acquired through their action on the fields. The transformations are the only ones that leave the gauge-field, but not the matter fields, invariant.\footnote{Indeed, such special configurations are part of the structure of the physical quotient of the configuration space of the theory by the gauge transformations, $\F/\G$: the quotient is not a manifold, but a stratified manifold, and the orbits whose fields are stabilized by subgroup of $\G$  are the `strata'.}
   That means that  in any background, malleable transformations that don't belong to these subgroups  would lack DES. That is, rigid symmetries \textit{may} have DES in a given special background and yet lose that significance  for a generic background. Such  subgroups  are ``plucked out'' from the surrounding  malleable group by satisfying certain equations, e.g. the Killing equations, which are themselves physically significant. 
 
 Such subgroups of rigid symmetries are usually called \textit{stabilizers}, and the configurations they stabilize are called \textit{reducible} (cf. footnotes \ref{ftnt:periodic} and \ref{ftnt:reducible}).  
 And the same concepts apply to gauge theories. In the Abelian case all $A\in \mathcal{A}$ are reducible, and they all possess the same stabilizers, namely, the constant gauge transformation. The non-Abelian case is much more similar to the spacetime case (see  \ref{sec:ambiguities} for a brief description): generic $A$ are \textit{not} reducible, but some are. Being reducible is a physically significant fact: in the quotient space $[\mathcal{A}]$, the orbits of reducible configurations are qualitatively different than the generic orbits.\footnote{Note that the notion of reducibility covaries with the notion of stabilizers. Namely, if a configuration $\varphi$ is reducible, with a given stabilizer $f\in \G$, then for any given $g\in \G$, $\varphi^g$ will also be stabilized by $gfg^{-1}$. This qualitative difference between the orbits renders  the quotient space into a \textit{stratified} manifold: i.e. a space formed by a concatenation of boundaries (see \cite{kondracki1983, Fischer} and, for a philosopher-friendly description of stabilizers, and their relation to conserved charges,  \cite[Secs. 3.3.5-7]{GomesStudies}) \label{ftnt:stab}}


\section{Conclusions}\label{sec:conclusions}

\subsection{Summary}\label{sec:conc_sum}
Broadly speaking, in this paper I have explored the role of GNSS  in the context of gauge theories.  I dissected the meaning and occurrence of  ``empirically significant subsystem symmetries'', whose existence and characteristics are still matters of debate in the philosophy of physics.  To make matters concrete, I have focused on Yang-Mills theory (and, in more detail, electromagnetism) and restricted subsystems to be demarcated by spacetime regions. 

In order to better adjudicate the debate between the opposing sides---represented by Greaves and Wallace (GW) on one side, and Brading and Brown (BB) on the other--- it was first necessary to clear the ground by  introducing new nomenclature. The standard nomenclature of local and global gauge symmetries is perfectly adequate if the system under study is the entire Universe, as is usually the case. But if one wants to discuss subsystems, and needs to distinguish between symmetries acting also at these different levels, the standard nomenclature becomes awkward. The awkwardness is apparent when one refers, e.g.: to `a global subsystem symmetry'. The new nomenclature uses four labels  to express two logically independent distinctions: namely (i) whether symmetries act only regionally or universally, and  (ii) whether they depend on an infinite or a finite set of parameters, i.e. whether they are rigid or malleable when acting throughout a spacetime region or throughout the universe. It also disentangles possible confusions with  `non-local and local' functions. 

With these definitions in place, and in the specific setting of Yang-Mills theory in bounded regions, I  found myself in agreement with GW  in their  criticism of  BB. Namely, BB prematurely dismissed the possibility of  DES tranformations  by assuming that  such a transformation would always create discontinuities in the boundary between the regions. The difference between GW and BB can be briefly summarized thus: one should be concerned with smooth composition of regional  gauge and matter fields (GW), as opposed to a smooth composition of gauge transformations (BB), which are not physical. 

But the gluing of the gauge and matter fields is also not physical; only the composition of their gauge equivalence classes is. This misunderstanding is reflected in GW's explicit assumption of supervenience on regions (cf. GSS' in Section \ref{sec:GWBB_error}), that is, that  regional states should uniquely define the universal state.  GW admit this condition  fails for  gauge theories in certain gauge-invariant bases (holonomies for the Abelian case), but it  even clearly fails for Galileo's ship.  Thus GSS seems inconsistent with an appropriate definition of   DES for gauge theories. 

My interpretation of  GW and BB's oversight is that they forget  aspects of the  Definition \ref{def:DES} of DES:  they do not check whether  condition $\mathit{1}$ is satisfied explicitly (by employing gauge-invariance information) and they do not  consider condition $\mathit{2}$ as involving physical information that is intrinsic to a subsystem or region.  


Having noticed these issues, we could summarize the journey from  the present results  to  GW and from there to BB as follows: one should be concerned with the smooth composition of the  physical data, as given by gauge-equivalence classes, and not with smooth composition of gauge and matter fields through  given representatives (GW), and much less with the smooth composition of gauge transformations (BB).  

Finding the criteria for smooth composition of the information contained in the equivalence classes $[\varphi_\pm]$ still presented a challenge.   The challenge was: can their physical content be composed into a universal physically valid state? 

The way we got around the problem of gluing physical content was to use an `externally sophisticated view of symmetries'.\footnote{Such a view is intimately related to composition, as  advocated by Rovelli \cite{RovelliGauge2013}; see also \cite{GomesStudies} for the same issue in a  context of regional subsystems.} Namely,  I employed `reduction', or better, `projection',  as a means to identify the regional and universal physical contents---each non-locally determined \textit{within} its corresponding domain,---and then I employed sophistication for gluing.

We found that for electromagnetism as coupled to a scalar Klein-Gordon field---when the Klein Gordon fields were taken to vanish at the boundary in between  the regions, but not in the bulk within them---there can be multiple universal physical states formed by gluing the same regional physical states. That is, a failure of uniqueness in the gluing  creates the gap  which gives a residual variety of universal physical states. In other words,  a variety of universal physical states can be built from the same regional physical states. These regional and universal  states fulfill our Definition \ref{def:reg_ins} for GNSS, and thereby also correspond to purely relational DES, as described in Definition \ref{def:DES}. 

Even though I have used `external sophistication', Theorem \ref{cor}, including its significance for DES, is completely compatible with viewing gauge degrees of freedom as ``descriptive fluff''.  The fact that the theory admits such a particular sort of redundancy is related to  the particular sort of non-locality of its gauge-independent degrees of freedom.  

In other words,  gauge-invariant quantities are to some extent non-local (L9 in Earman's classificatory scheme \cite{Earman_local}), which means there is in principle ``room  to explore'' between the whole and the sum of the parts. And it turns out that the known aspects of non-locality in gauge theory match the non-local aspects of GNSS.

One can see this is as follows. Gauss constraints are the defining characteristic of gauge theories: they stipulate that charges couple to fields in such a way that their conservation laws are dynamically respected. But Gauss's law can clearly fluster cluster decomposition: once one knows the electric flux around a closed surface, one knows precisely the amount of charge within it. Quantities measured on the totality of the boundary are therefore not independent of other quantities measured in the bulk. Of course, this type of ``synchronic non-locality'' is not causal; it only represents  ``non-locally possessed'' properties (in the language of \cite{Belot1998, Healey_book}). That is, there should be room for \textit{holism}, which this paper has drawn on. 
 
 The work of \cite{GomesRiello_new} reported here precisely delineates the sort of non-locality involved.  For generic regional states in  Yang-Mills theory, one does \textit{not} have   GNSS: in most circumstances we can describe the physical state of the whole by describing the physical state of its composing regions. When the regional states do not uniquely determine the universal state, GNSS ensues; and with it, DES as I have defined it. That is, it is a particular non-locality that is responsible for the gap between the regional gauge-invariant information and the universal gauge-invariant information; it is  this gap from which DES emerges. 
 
 Moreover, as expected,  the group of symmetries with DES has intimate connections with the charge group of the theory: For both Abelian and non-Abelian theories  the residual variety of universal physical states can be parametrized by sub-algebras of $\mathrm{Lie}(G)$. And these rigid symmetries with DES are in 1-1 correspondence with the possible (covariantly) conserved charges in the region (cf.  \cite[Sec. 4.3.2]{GomesRiello_new}). In the Galileo's ship example, the variety is given by the action of the Galilean group, `$\text{Boosts}\ltimes\text{Euclidean}$,' on the subsystem, and the respective charges are the conserved momenta. 
 
What these findings suggest is that (for simply connected regions,\footnote{For both Abelian and non-Abelian Yang-Mills,  for non-simply connected manifolds we get something beyond $G$: namely, we also get the first homology group, whose dimension is finite and given by the so-called first (equivariant) Betti number. Such a contribution from non-trivial topology was to be expected from the source of our results: holism.} and at the perturbative level in the case of the non-Abelian theory) the only failure of holism of gauge theories is precisely that related to the total covariantly conserved charges within any region. That is, the only failure is due to the Gauss law.

 In the landscape of the debate, we thus locate ourselves somewhere  between the two opposing views. On the side of BB and the orthodoxy,\footnote{In fact, in this point, we roughtly align with Kosso \cite{Kosso}---but not always: only when covariantly conserved charges exist---and not with Brown and Brading, who want to claim that even rigid internal symmetries have no DES.}  we find that indeed only rigid, but no malleable symmetries may have direct empirical significance in gauge theory. In fact, surprisingly, the rigid ones   correspond to  `the global'  gauge transformations, $G$ (for non-Abelian: only for very particular perturbed configurations, that can carry associated covariantly conserved charges). On the side of GW, we find that gauge theories may indeed harbor symmetries that have direct empirical significance. But  to force  my conclusion into either pigeonhole, or even a combination,  would be to shave off some of the important subtleties of this situation.
 
In a few words: for the physical, i.e. gauge-invariant, content of the gauge fields, the whole can be more than the sum of the parts. Their difference manifests itself in a direct empirical significance of  (subgroups of) the charge group, and are in  1-1 relation to the conserved charges of the theory.  The same can be said for Galileo's ship scenario, where the difference manifests itself in the direct empirical significance of the Galilean group.

\subsection{The role of the subsystem}\label{sec:epistemic}

 The  conclusion we have arrived at says that DES are not inextricable from other effects of symmetries. In particular, I have related direct empirical significance (DES) to a failure of supervenience on subsystems (GNSS), and then stated  that GNSS is related to the indirect significance of symmetry (labeled IES), like conservation of charges. 
 
Given the sort  of holism we have seen in GNSS, it thus might be conceptually preferable to use global information about charges and other superselected
quantities pertaining to IES to parametrize the possible gluings of regions in a gauge-free manner. In that case, we may not want to invoke external sophistication. 

Certainly in practice we will
often lack the relevant global information, and so it makes sense to use
a local gauge-fixed parametrization to do calculations, make predictions, etc. Similarly, insofar as we only have a perturbative handle on
the gluing procedure in the non-Abelian case, and our current perturbation theory relies on compactly localized fields,  there is reason
to use gauge-fixed compactly-localized fields instead of gauge-invariant
non-compactly localized fields, and external sophistication gives a conceptual underpinning to the ensuing gluing procedures. But these sorts of concerns are probably best seen as
practical or technical barriers, not as in-principle reasons to adopt external
sophistication.

Indeed, from a global perspective external sophistication loses its warrant. But  then, so does the empirical relation between conserved charges and rigid symmetries, since DES necessarily involves an external perspective. 

\subsection*{Acknowledgements}

I would like to thank Aldo Riello, my collaborator in all things gauge-related, for many fruitful and insightful discussions and for co-developing the entire technical apparatus employed in this analysis. And similarly, I would like to thank Jeremy Butterfield, for many discussions, for many requests for clarification,  for patiently guiding me in all things philosophy-related, and for giving immensely valuable and thorough feedback on this paper.

\begin{appendix}

\section*{APPENDIX}

\section{Non-Abelian Yang-Mills}\label{app:nonYM}
\subsection{Notation for Yang-Mills theory coupled to matter}

The general case works with a  finite-dimensional \textit{charge group}, e.g. $G=\SU(N)$, with Lie algebra $\mathfrak{g}:=\mathrm{Lie}(G)$, e.g. $\mathfrak{g}=\mathfrak{su}(N)$. Given the charge group we define the  group of gauge transformations $\G=C^\infty(\Sigma, G)$, with composition given by pointwise action of $G$, i.e. $(g g')(x):=g(x)g'(x)$, and the respective infinitesimal version, $\fG= C^\infty(\Sigma,  \mathfrak{g})$ (with pointwise Lie algebra commutator). An element of $\G$ is a map: $g(\cdot):\Sigma\ni x\mapsto g(x)\in G$. The gauge fields and its gauge-transformed version $A^g$ are given by Lie-algebra valued space(time) 1-forms.  

We define, for $\d x^\mu$ a basis of 1-forms over $\Sigma$ and $\tau_\alpha$ a basis of the Lie algebra $\mathfrak{g}$
 \be\label{eq:gt}A=A^\alpha_\mu \d x^\mu \tau_\alpha\in \Lambda^1(\Sigma,\mathfrak{g})\quad \text{and}\quad A^g:=g^{-1}Ag+g^{-1}\pp g\ee
Infinitesimally, i.e. for an infinitesimal gauge-transformation $\xi\in \fG$, the gauge field transforms as 
\be\label{eq:gt_alg}A\mapsto A+\D \xi \quad \text{where}\quad \D\xi:=\pp \xi +[A,\xi].\ee
For full generality we can introduce charged fermions  in a fundamental representation: so for some vector space $W$, the 4-component Dirac spinor field (i.e. in $\bb C^4$) as:
 $$\psi \in C^\infty(\Sigma,\C^4\otimes W)\quad\text{and}\quad\psi^g=g\psi.$$
 And I will write the joint configuration as 
 $$ \varphi=(A, \psi)
 $$	 
	  I will denote the regional, unquotiented configuration spaces of each field sector (gauge field and matter, respectively) as  $\mathcal{A}_\pm=\{A_\pm \in \Lambda^1(\Sigma_\pm, \mathrm{Lie}(G))\}$, $\Psi_\pm=\{\psi_\pm\in C^\infty(\Sigma_\pm,\C^4\otimes W)\}$.  And, for the joint configuration spaces,  I write $\Phi_\pm=\{\varphi_\pm:=(A_\pm,\psi_\pm)\in \mathcal{A}_\pm\times \Psi_\pm\}$, writing $\Phi, \Psi, \mathcal{A}$.  I will omit the subscript $\pm$ for the corresponding universal configuration spaces. The restricted groups of gauge transformations will be denoted in analogous fashion: $\G_\pm=C^\infty(\Sigma_\pm, G)$, and all abstract quotient spaces are denoted by the square brackets, as in $[\Phi^\pm]:=\Phi^\pm/\G_\pm$,  and $[\F]:=\F/\G$.

 \subsection{Gluing for non-Abelian Yang-Mills theory}\label{app:gluing_YM}

  In the non-Abelian case we have the equations analogous to \eqref{eq:hor_conds} and \eqref{eq:Landau}, namely:
 \be\begin{cases}
  \D^i \delta A^\pm_i &=0 \\
   s^i \delta A^\pm_i{}_{|S}& =0\end{cases} \label{eq:hor_conds_YM}\ee
    \be\label{eq:Landau_YM}
\D^i \delta A_i =0\ee
And we use apply them to establish gluing through the analogous to \eqref{eq:gluing_simple}
 \be \label{eq:gluing_simple_YM}
\delta A:= (\delta A^++\D\xi^+)\Theta_++(\delta A^-+\D\xi^-)
\ee
  We obtain the following conditions on $\xi_\pm$:
  $$\begin{cases}
\D^2\xi_\pm=0\qquad \text{on}\qquad \Sigma_\pm\\
s_i\D^i(\xi_+-\xi_-)=0 \qquad \text{on}\qquad S\\
(\xi_+-\xi_-){}_{|S}=(\D^2_S)^{-1}(\D^c_S(   \delta A_c^+ -  \delta A_c^- )_{|S})\quad \text{on}\quad S
\end{cases}$$
(subscript $S$ means intrinsic to the surface).
  
  We can solve these emerging conditions (see Section  4 in \cite{GomesRiello_new}), obtaining:
\begin{align} \label{eq:gluing_nonabelian}
\xi^\pm = \zeta_{(\pm)}^{ \pm\Pi}\quad\text{with}\quad \Pi=\Big(\mathcal{R}^{-1}_+  + \mathcal{R}^{-1}_-\Big)^{-1}\left(  (\D^2_S)^{-1}\D^c_S(   \delta A_c^+ -  \delta A_c^- )_{|S}\right)
\end{align}
where, in each region,  $\zeta^\phi$ stands for a covariantly harmonic function, satisfying $\D^2\zeta=0$, with Neumann boundary conditions at $S$ given by $s^i\D_i\zeta{}_{|S}=\phi$, and where the subscript $S$ denotes ``intrinsic to  $S$'', and the intrinsic coordinates to $S$ are given by $c$, and where $\mathcal{R}$ is the so-called Dirichlet-to-Neumman operator. Briefly,   $\mathcal R$ functions as follows: a given harmonic function with Dirichlet conditions---these conditions are the input of $\mathcal R$---will possess  a certain normal derivative at the boundary; i.e. will induce  certain Neumann conditions there---these conditions are the output of $\mathcal R$. That is, let $\zeta^u$  be a  harmonic function with Neumann boundary condition, then for $\zeta_u$ a covariantly harmonic function with Dirichlet boundary condition, $(\zeta_u)_{|S}= u$, the Dirichlet to Neumann operator $\mathcal{R}$ is defined as $\zeta_u=\zeta^{\mathcal{R}(u)}$ (i.e. it finds the harmonic function with Neumann condition corresponding to one with a  Dirichlet condition).
 
   \subsection{Non-Abelian GNSS}\label{sec:ambiguities} 
   There  are two sources of under-determination of solutions to \eqref{eq:gluing_simple_YM}: one topological and one from homogeneous fields.   The first arises from the possibly non-trivial  first (equivariant) homology group of $\Sigma$, we look at the first in Section  \ref{sec:topo_hol} and at the second in Section  \ref{sec:stab_YM}. 
   
    \subsubsection{Topological holism}\label{sec:topo_hol}
      We may have a universal field $\delta A$ satifying \eqref{eq:Landau_YM} which, when restricted to each $\Sigma_\pm$ is pure gauge, i.e. of the form $\D\xi_\pm$. In other words,   the room for  discrepancy is equivalent to broken homology cycles. This is easier to see in the Abelian case, where $\delta A$ would be a one form which is exact in the simply connected patches, but only closed in the entire manifold (see \cite{GomesRiello_new}, Section  4.7 for an example). By the  Poincar\'e lemma, this occurs if and only if the topology of the manifold is non-trivial.  In other words, universal physical processes in this case may not come from regional physical processes; some physical processes are universal/global in nature. 
  
  But of course for topologically non-trivial manifolds we should expect the whole to contain more information than the sum of the parts, foiling \textit{Antiholism}! The arising GNSS would contribute with the suitable homology group  to  the variety set $I$ in \eqref{eq:reg_ins}. Such a contribution has a finite number of generators (given by the first Betti number, i.e. the rank of the first homology group).\footnote{This topological fact should be consequential to the labeling of the inequivalent representations of the $\theta$ vacua (see \cite{Strocchi_phil}).}

    \subsubsection{Stabilizer ambiguity}\label{sec:stab_YM}
     Barring non-trivial topology, is there another source of under-determination of $\xi^\pm$ also in the non-Abelian case?   
 To arrive at an answer, note that we are using the gauge-fields as a reference. It is through them that we fix the $\xi$ and   parametrize the  reduced/quotient configuration space $[\F]=\F/\G$.  Therefore one source of GNSS would be a possible under-determination of $\xi_\pm$.
  
   This under-determination occurs if and only if there exist $\chi_\pm\in \fG$ such that
\be \D\chi_\pm=\pp \chi_\pm+[A_\pm,\chi_\pm]=0
\ee
for $A=A_+\Theta_++A_-\Theta_-$ being  the smooth field configuration around which we are considering perturbations.\footnote{This assumption is not crucial, since the way we implement the perturbative gauge fixing is covariant with respect to gauge transformations in $\mathcal{A}$. } This is, again, the ``Killing'' (or infinitesimal  stabilizer) equation for gauge transformations.  Namely, under-determination of $\xi_\pm$ occurs if and only if there exist $\chi_\pm$ which \textit{stabilize} the entire $A_\pm$, in which case we could use either $\xi_\pm$ or $\xi_\pm+\chi_\pm$ for gluing.

Such stabilizers are generalizations of the constant potential shifts in electromagnetism. Here the entire regional field must be stabilized by $\chi_\pm$, and this singles out from $\G^+$ the  regional elements which can potentially exhibit DES. 

As with Killing directions for spacetime geometries, such elements are  hard to come by: they're generically trivial (for non-Abelian charge groups $G$), and, when non-trivial, they are generated by a finite-dimensional basis. 

In sum: there may still be some small variety in $\xi^\pm$, which is innocuous as far as the gauge field is concerned. However, as we will shortly see, this variety  can be relationally felt by regional matter fields.

        \subsubsection{Obtaining variety from the stabilizer ambiguity}\label{sec:stab_amb}

 When we include matter fields, the universal field is composed as:
 \be\label{eq:gluing_matter} \delta \psi= (\delta \psi^++\xi^+\psi_+)\Theta_++(\delta \psi^-+\xi^-\psi)\ee 
 Since we are not using the matter fields to parametrize the quotient/moduli/reduced configuration spaces, there are no analogues of equations \eqref{eq:hor_conds} and \eqref{eq:Landau} to be implemented for it. But smoothness of the field does impose certain conditions. In fact,   different choices of stabilizer, adding $\chi_+\neq \chi_-$ to $\xi_+, \xi_-$, respectively, will render the matter fields incompatible at the boundary.\footnote{  That is, unless $\chi^\pm$ also stabilize $\psi^\pm$. 
 For non-Abelian fields, there may still be internal stabilized directions, i.e. for particular configurations $\tilde\psi_\pm\neq 0$, there may be $\tilde\chi_\pm\neq 0$ such that $\tilde\chi_\pm\psi_\pm=0$. This is not true for U(1),  there  $\tilde\chi_\pm\psi_\pm\neq 0$ for any non-vanishing gauge transformation and matter field. But I do not know of any non-trivial, shared stabilizers for both fields in such a situation. }
 
 Therefore, if the matter fields do \textit{not} vanish at the boundary, we can assume there is again regional determinacy as per Definition \ref{def:reg_ins}. So we assume they vanish there; a condition postulated by GW  as ``dynamical isolation'' of the two regions.  And also,  if the matter field vanishes not only at the boundary, but  everywhere,  no GNSS, i.e. GSS arises, yet again.      Thus, I assume: \textit{the matter fields vanish at the boundary}, \textit{but not in the bulk} of the regions.  
 
To illustrate the precise emergence of the Lie algebra $\mathfrak{g}$, we  resort to a `best-case' scenario. That is, apart from the conditions stipulated above for the matter fields, we will take the gauge field configuration around which we are perturbing to be the `vacuum', $A=0$.  In that case, the gauge covariant differential just becomes the standard differential (as in the Abelian, electromagnetic case), i.e. $\D\rightarrow \pp$, and the stabilizer equation becomes $(\pp \chi^\alpha(x))\tau_\alpha=0$.  Then not only is  the space of stabilizers $\chi$  a finite-dimensional vector space, closed under commutation, but for this case it forms a Lie algebra homomorphic to $\mathfrak{g}$.  

Then we have a parametrized solution
  \be\label{eq:gluing_matter} \delta \psi_{(\chi^+, \chi^-)}= (\delta \psi^++(\xi^++\chi^+)\psi_+)\Theta_++(\delta \psi^-+(\xi^-+\chi^-)\psi)\ee 
 But we are still not done. For $\chi^+= \chi^-$, $\delta \psi$ is still universally gauge equivalent to $\delta \psi_{(\chi^+, \chi^-)}$, i.e. they differ by a universal, rigid gauge transformation. Therefore, what really matters for universal variety is the \textit{difference}: $\chi_+-\chi_-$. This difference is generated by just one copy of $\mathfrak{g}$. That is, given the $\delta A_\pm, \delta \psi_\pm$, we have:
\be\label{eq:full_DES} (\delta A, \delta \psi_{(i)})=(\delta A^++\D\xi_{(i)}^+, \delta \psi^++\xi^+_{(i)}\psi^+)\Theta_++(\delta A^-+\D\xi^-, \delta \psi^-+\xi^-\psi^-)\Theta_-\ee
where ${\xi^+_{(i)}}(x):=(\xi^+_\alpha(x)+i_\alpha )\tau^\alpha$, for spacetime constant coefficients $i^\alpha$, with $\tau_\alpha$  a basis of $\mathfrak{g}$ (which also implies $\D\xi^+=\D\xi^+_{(i)}$), i.e. the $i$ parametrize the Lie algebra of the charge group $G$. 
 
  Finally, we have arrived at our destination: I have shown that the variety set $I$, at least infinitesimally and in the best-case scenario,  can in some sense recover the charge group $G$, i.e. I have characterized the infinitesimal version of \eqref{eq:gluing_quotients}, with:
    \be    [\delta \varphi^{(i)}]=[\delta\varphi^+]\cup^S_{(i)} [\delta\varphi^-],\quad i\in  \mathfrak{g},\quad\text{with}\quad  [\delta\varphi^{(i)}]\neq  [\delta\varphi^{(i')}]\quad\text{iff}\quad i\neq i'\ee
    For non-Abelian groups, it is not clear how to obtain the finite version of this equation, replacing $\delta \varphi$ with $\varphi$, i.e. with 
        \be    [ \varphi^{(i)}]=[\varphi^+]\cup^S_{(i)} [\varphi^-],\qquad i\in I \simeq G\ee
        But in the Abelian case, the tools utilized here \textit{do} allow an integration to the finite setting.\footnote{This is related to the `dressing formalism'', see Section  9 of  \cite{GomesHopfRiello} and \cite{Francois, Attard2018}.} 
  
  Moving away from the best-case scenario of $A=0$: to obtain GNSS the base configuration $A$ must still  be somewhat homogenous; it must admit internal directions which leave it unchanged. But as mentioned above, a general theorem about Killing fields (and stabilizers), shows that Killing directions are always generated by a finite dimensional basis; in the case of internal gauge transformations, this basis consists of at most $\textrm{dim}(\mathfrak{g})$ elements, closed under commutation, etc. Thus even away from the best-case, we obtain $I$'s isomorphic to sub-algebras of $\mathfrak{g}$ (we would also have to replace $\tau_\alpha$  in the definition of $\delta\psi_{(i)}$ for a choice of basis $\{\chi_\alpha\}$ of the appropriate Killing, or stabilizing fields).

Note that, as had to be the case,  according to Definition \ref{def:reg_ins}, there exists a \textit{regional variety}  implicit in equation \eqref{eq:full_DES}, and this variety recovers also the definition of DES through transformations, if  the transformations referred to  in Definition \ref{def:DES} are  taken to be e.g.: $(\delta A, \delta \psi)\rightarrow (\delta A, \delta \psi_{(i)})$.

\end{appendix}


\end{document}